\documentclass[sn-aps]{sn-jnl}%

\jyear{2021}%

\theoremstyle{thmstyleone}%

\theoremstyle{thmstyletwo}%

\theoremstyle{thmstylethree}%

\begin{document}
\title[Grey area in Embedded WMLES on a nacelle-aircraft configuration]{Grey area in Embedded WMLES on a transonic nacelle-aircraft configuration}

\author*[1]{\fnm{Marius} \sur{Herr}}\email{m.herr@tu-bs.de}

\author[2]{\fnm{Axel} \sur{Probst}}\email{axel.probst@dlr.de}

\author[1]{\fnm{Rolf} \sur{Radespiel}}\email{r.radespiel@tu-bs.de}

\affil*[1]{\orgdiv{Institute of Fluid Mechanics}, \orgname{TU Braunschweig}, \orgaddress{\street{Hermann-Blenk-Str. 37}, \city{Braunschweig}, \postcode{38108}, \state{Lower Saxony}, \country{Germany}}}

\affil[2]{\orgdiv{Institute for Aerodynamics and Flow Technology}, \orgname{DLR}, \orgaddress{\street{Bunsenstr. 10}, \city{Göttingen}, \postcode{37073}, \state{Lower Saxony}, \country{Germany}}}

\abstract{
A scale resolving hybrid RANS-LES technique is applied to an aircraft - nacelle configuration under transonic flow conditions using the unstructured, compressible TAU solver. 
Therefore a wall modelled LES methodology is locally applied to the nacelle lower surface in order to examine shock induced separation. In this context a synthetic turbulence generator (STG) is used to shorten the adaption region at the RANS – LES interface. Prior to the actual examinations, fundamental features of the simulation technique are validated by simulations of decaying isotropic turbulence as well as a flat plate flow.
For the aircraft - nacelle configuration at a Reynolds number of 3.3 million a sophisticated mesh with 420 million points was designed which refines 32\,\% of the outer casing surface of the nacelle. The results show a development of a well resolved turbulent boundary layer with a broad spectrum of turbulent scales which demonstrates the applicability of the mesh and method for aircraft configurations. Furthermore, the necessity of a low dissipation low dispersion scheme is demonstrated.
However, the distinct adaption region downstream of the STG limits the employment of the method in case of shock buffet for the given flow conditions.
}

\keywords{hybrid RANS-LES, wall-modelled LES, synthetic turbulence, aircraft configuration, transonic flow, shock induced separation}

\maketitle
 \section{Introduction}

Transonic flows about aircraft configurations exhibit complex, instationary flow phenomena such as oscillating shock fronts with boundary layer separation. 
This so-called buffet phenomenon causes unsteady aerodynamic loads which might endanger the flight safety.
Therefore a fundamental understanding of the related flow physics is of particular interest to be able to find specific technical solutions which control this phenomenon.
The present study examines a XRF-1 aircraft model which represents a wide-body long-range configuration and was designed by Airbus.
An Ultra High Bypass Ratio (UHBR) nacelle is coupled to the model which represents a modern and efficient jet engine that is modelled as flow-through nacelle for wind tunnel testing.
Due to the large circumference of the nacelle, a close coupling by means of a pylon to the wing lower side is necessary. This channel-like arrangement of nacelle, pylon, wing and fuselage causes the development of an accelerated flow which triggers the formation of transonic shocks within this area. Depending on the exact flow conditions these shocks evolve into buffet with significant loads.
Initial investigations in the framework of the DFG (Deutsche Forschungsgemeinschaft) funded research group have shown a complex system of shock fronts \cite{spinner2021design}. As a first step toward representing  this complex system with a sophisticated numerical method this study focuses on a single shock front located at the lower side of the nacelle.

Numerous numerical investigations have investigated the problem of buffet onset with well established unsteady Reynolds-averaged Navier-Stokes (URANS) methods. However, it is well known that even highly developed Reynolds stress based URANS models show deficiencies in describing the dynamics of separated boundary layer as well as the aerodynamic effects of large flow separations \cite{cecora2015differential}.
Also, due to high, flight relevant Reynolds numbers a broad scale of turbulent structures arise for the given flow phenomenon.
Therefore a simulation technique that provide both high spatial and temporal resolution is required.
Direct Numerical Simulation (DNS) resolves all turbulent scales but is so far restricted to simple geometries at low Reynolds numbers due to its unfeasible computational effort for flight relevant flows.
Therefore a Large Eddy Simulation (LES) technique is required which only resolves large turbulent scales whereas small, isotropic scales are modelled. Since an application of LES to the entire aircraft configuration is still computationally too expensive a hybrid RANS - LES technique is employed.
In the present study the wall modelled LES (WMLES) method within the Improved Delayed Detached Eddy Simulation (IDDES) methodology is used \cite{shur2008hybrid}. Depending on the spatial discretisation, up to $5\,\%$ of the wall adjacent boundary layer is modelled by the RANS equations. Additionally, the area of WMLES is embedded around the transonic shock such that all relevant flow areas are enclosed. This corresponds to 32\,\% of the outer casing surface of the nacelle. The remaining flow field of wing, body, pylon and nacelle is modelled with a URANS model.
The embedded WMLES (EWMLES) requires an injection of synthetic turbulence at the RANS-LES interface which is located at the leading edge of the nacelle for the present configuration. Otherwise, a so-called grey area would arise which describes a region of underresolved turbulence directly downstream of the RANS-LES boundary. To this end the synthetic turbulence generator (STG) devised by \cite{Travin2002} is employed. Nevertheless, using this method, a transitional region from modelled to fully resolved turbulence is still present and is referred to as adaption region in this study. The analysis of this adaption region with regard to its length and behaviour of relevant flow quantities in this area are of major interest.  
Thus, especially the transient establishment of resolved turbulence within the WMLES area and the fundamental applicability of the method to the aircraft configuration are the focus of this study.

The study is structured as follows. The employed WMLES model in conjunction with the STG is described in detail in subsection \ref{subsec:hrlm} and \ref{subsec:stg}, respectively. Subsequently a thorough description of the employed low dissipation low dispersion (LD2) numerical scheme is given in \ref{subsec:LD2}. The following section \ref{sec:basic_validation} provides a basic validation of the Embedded WMLES based on the SST-RANS model by means of flows of decaying isotropic turbulence and a flow about a flat plate. 
The results of the application to the XRF-1 configuration are presented in section \ref{sec:results}. An extensive description of the mesh design with regard to the extension of the WMLES area, the used refinement criteria and its application to the actual mesh environment are presented (Sec. \ref{sec:grid_design}).
Results of the transient WMLES establishment are then shown and assessed in section \ref{sec:transient_process}. The analysis of temporally and spatially averaged flow quantities in the area related to the STG is carried out (Sec. \ref{sec:investigation_of_grey_area}).
Finally, sensitivity studies with regard to the position of the RANS-LES boundary (Sec. \ref{sec:results_positioning_STG}) and the effect of using a standard numerical scheme instead of the low dissipation scheme (Sec. \ref{sec:results_numerical_scheme}) is presented. This paper is closed by a final summary of all research findings (Sec. \ref{sec:conclusions}).

\section{Numerical Methods}\label{sec:num_meth}
The flow simulations in this paper use the unstructured compressible DLR-TAU code \cite{Schwamborn2006} which numerically solves the flow and model equations on mixed-element grids (e.g. hexahedra, tetrahedra, prims) via the finite-volume approach.
It applies $2^{nd}$-order discretization schemes for both space and time, together with low-Mach-number preconditioning for flows that are close to the incompressible limit.
Implicit dual-time stepping allows adapting the time step in unsteady simulation to the physical requirements (i.e. related to the convective CFL-criterion), avoiding numerical stability restrictions.

The relevant methods for embedded wall-modelled LES, i.e. the overall (hybrid) turbulence model, the method to generate and inject synthetic turbulence and the required local adaptation of the numerical scheme, are outlined in the following.

\subsection{Hybrid RANS-LES Model}\label{subsec:hrlm}
The present embedded wall-modelled LES approach relies on the Improved Delayed Detached-Eddy Simulation (IDDES) \cite{shur2008hybrid} which combines local RANS, DES (i.e. RANS-LES) and wall-modelled LES (WMLES) functionalities in a seamless, automatic manner.
This is achieved by a single \emph{hybrid} length scale replacing the integral turbulent scale $l_{\mbox{\tiny RANS}}$ in the underlying RANS model, which is the two-equation SST model \cite{Menter1994} in the present work.
The hybrid length scale reads:
\begin{equation}\label{eqn:lhyb_IDDES}
l_{hyb}  = \tilde f_d \left( {1 + f_e } \right)l_{\mbox{\tiny RANS}}  + \left( {1 - \tilde f_d } \right)l_{\mbox{\tiny LES}}  \quad .
\end{equation} 
Here, the function $\tilde f_d = \max \left\{\left(1 - f_{dt} \right), f_B \right\}$ is the main blending switch between the different modelling modes, where $f_{dt}$ and $f_B$ depend on local grid and flow properties (cf. \cite{shur2008hybrid}).

In WMLES mode ($f_{dt} \equiv 1$ and, thus, $\tilde f_d \equiv f_B$), if resolved turbulent content enters an attached boundary layer, a RANS layer is kept near the wall and sized according to the local grid resolution, thus circumventing the extreme grid requirements of wall-resolved LES at high Reynolds numbers. 
However, since no wall-functions are applied in the present work, the equations need to be solved down to the wall with a (normalized) near-wall grid spacing of $y^+(1) \leq 1$.
The additional \emph{elevating} function $f_e$ is designed to reduce the well-known log-layer mismatch in WMLES.

In the largest (outer) parts of the boundary layer, $l_{hyb} \equiv l_{\mbox{\tiny LES}} = C_{\mbox{\tiny DES}} \Delta$, which approximates the behaviour of a Smagorinsky-type sub-grid model for LES.
The model constant $C_{\mbox{\tiny DES}}$ is usually calibrated for canonical turbulent flow, such as decaying isotropic turbulence (DIT), see Sec.~\ref{sec:dit}.
However, since wall-bounded flows typically require a different calibration than free turbulence, another modification compared to standard DES/LES is introduced in the filter width $\Delta$:
\begin{equation}\label{eqn:delta_iddes}
\Delta = \Delta_{\mbox{\tiny IDDES}}  = \min \left\{ {\max \left[ {C_w \cdot d_w , C_w \cdot h_{\max }, h_{wn} } \right], \Delta_{\mbox{\tiny DES}} } \right\} \quad ,
\end{equation}
where $C_w = 0.15$. 
In essence, this near-wall limitation of the filter width compensates for this flow-type dependency and allows using a unique $C_{\mbox{\tiny DES}}$ value for both wall-bounded and off-wall turbulent flow.
More details on this modification are found in \cite{shur2008hybrid}.\\

For embedded WMLES, the IDDES in TAU can be locally forced to WMLES mode according to external user input, e.g. inside boxes or other suitable geometric sub-areas of the flow domain. 
This is achieved by setting the function $f_{dt}$ to 1 downstream of the desired RANS-WMLES interface, thus safely  reducing the eddy viscosity from RANS to WMLES level \cite{probst2017evaluation}.

\subsection{Synthetic Turbulence Generation}\label{subsec:stg}

In this work, synthetic turbulent fluctuations at the streamwise RANS-LES interface are provided by the Synthetic Turbulence Generator (STG) of Adamian and Travin \cite{Adamian2011a} with extensions for volumetric forcing by Francois \cite{francois2015forced}.
This STG generates local velocity fluctuations from a superimposed set of $N$  Fourier modes as:
\begin{equation}\label{eq:STG_fluc}
 \vec{u}'_{ST} 
 = \vec{\underline{A}} \cdot \sqrt{6}\sum_{n=1}^N\sqrt{q^n}\left[\vec{\sigma}^n\cos\left(k^n\vec{d^n}\cdot\vec{r}'+\phi^n+s^n\frac{t'}{\tau}\right)\right] \quad ,
\end{equation}
where the direction vectors $\vec{d}^n$ and $\vec{\sigma}^n \perp \vec{d}^n$, the mode phase $\phi^n$, and the mode frequency $s^n$ are randomly distributed.
A realistic spectral energy distribution of the mode amplitudes $q^n$ is achieved by constructing a von K\'{a}rm\'{a}n model spectrum from RANS input data and a local grid cut-off.
The RANS data, which is automatically extracted from just upstream the RANS/LES interface, is also used to scale the fluctuations via the Cholesky-decomposed RANS Reynolds-stress tensor $\vec{\underline{A}}$.

For realistic temporal correlations in a volumetric forcing domain, the position vector $\vec{r'}$ and the time $t'$ are modified in accordance with Taylor's frozen velocity hypothesis, see \cite{francois2015forced} for details.

\paragraph*{Synthetic-Turbulence Injection}
To inject the synthetic fluctuations from Eq.~(\ref{eq:STG_fluc}), a forcing volume 
with a streamwise extent of about half the local boundary-layer thickness is marked
just downstream of the RANS/LES interface.
Inside this volume, a momentum source term is added \cite{Probst2020} which approximates the partial time derivative of the synthetic fluctuations
as:
\begin{equation}\label{eq:forcing_source}
 \vec { Q } = \frac{\partial\left(\rho \vec{u}'_{ST}\right) }{\partial t}  \approx \frac { 3\left( {\rho \vec{u}'_{ST} } - \rho \vec{u}'^n \right) -\left( \rho \vec{u}'^n - \rho \vec{u}'^{n-1} \right)  }{ 2\Delta t }  \ .
\end{equation}
This discretization corresponds to the $2^{nd}$-order backward difference scheme used for unsteady simulations with TAU.
By computing the fluctuation values of the previous time steps from the actual flow field, i.e. as $\vec {u' } ^{ n } = \vec {u} ^{ n } -  \langle \vec {u} \rangle $ and $\vec { u' } ^{ n-1 } = \vec {u} ^{ n-1 } -  \langle \vec {u} \rangle $, 
the synthetic target field (Eq.~\ref{eq:STG_fluc}) can be reproduced rather accurately in the simulation, even though running time averages are required.
An additional Gauss-like blending function with a maximum value of 1 around the streamwise center of the forcing volume is multiplied to the source term in order to prevent abrupt variation of the forcing.

\subsection{Hybrid Low-Dissipation Low-Dispersion Scheme}\label{subsec:LD2}
Since scale-resolving simulation methods like IDDES involve explicit modelling of the sub-grid stresses, the overall accuracy relies on low spatial discretization errors in the LES regions of a given grid.
Concerning resolved turbulence, there are two types of error that mainly stem from the discretized convection of momentum: while numerical dissipation damps the turbulent fluctuations and would lead to under-predicted Reynolds stress, numerical dispersion distorts the shape of resolved turbulent structures.

For that reason, the present simulations apply a hybrid low-dissipation low-dispersion scheme (HLD2) \cite{Probst2016}, which combines different techniques to optimize the convection scheme for local scale-resolving simulations using unstructured finite-volume solvers.

To provide low numerical dissipation, the spatial fluxes are calculated from Kok's \cite{kok2009high} skew-symmetric central convection operator, which allows for kinetic-energy conservation (i.e., it is non-dissipative) on curvilinear grids in the incompressible limit. 
For compressible flow on general unstructured grids, a classic blend of 2nd- / 4th-order artificial matrix-dissipation is added to ensure stability around shocks and in smooth flow regions.
Compared to RANS computations, however, the 4th-order dissipation has been strongly reduced by manually optimizing its parameters in LES computations of the channel flow, yielding e.g. a global scaling factor of $\kappa^{(4)} = 1/1024$ and a reduced Mach-number cut-off in the low-Mach-number preconditioning matrix. 

Moreover, to minimize the dispersion error of the second-order scheme, the skew-symmetric central fluxes are based on linearly-reconstructed face values $\phi_{L, ij}$, $\phi_{R, ij}$ using the local Green-Gauss gradients $\nabla_0 { \phi }$.
Exemplarily, a generic central flux term reads:
\begin{equation}\label{eqn:ld2_flux}
\phi_{ij, \alpha} = \frac{1}{2} \left( \phi_{L, ij} + \phi_{R, ij} \right) =  \frac{1}{2} \left( \phi_i + \phi_j \right) + \frac{1}{2} \alpha \left( \nabla_0  { \phi_i} - \nabla_0 { \phi_j } \right) \cdot \mathbf{d}_{ij}  \quad ,
\end{equation}
where $\mathbf{d}_{ij}$ is the distance between the points $i$ and $j$.
With an extrapolation parameter of $\alpha = 0.36$ the scheme was found to minimize the required points per wavelength for achieving a given error level in a 1-D wave problem, see \cite{Loewe2016} for details.

\paragraph*{Blended Scheme for Hybrid RANS-LES}
While the low-error properties of the LD2 scheme are essential for accurate LES and WMLES predictions with TAU \cite{Probst2016}, the pure RANS and outer flow regions in hybrid RANS-LES are less dependent on such numerical accuracy.
Moreoever, although the LD2 scheme has been globally applied in hybrid RANS-LES, complex geometries like the present XRF-1 configuration and corresponding unstructured grids may induce local numerical instabilities that are not damped by low-dissipative schemes.
For this reason, we apply the LD2 scheme in a hybrid form \cite{Probst2016} where all parameters of the spatial scheme, $\Psi_i$, are locally computed from a blending formula: 
\begin{equation}\label{eq:num_blending}
\Psi_i = (1 - \sigma) \cdot \Psi_{i, \mbox{\tiny LD2}} + \sigma \cdot \Psi_{i, \mbox{\tiny Ref}} \quad .
\end{equation}
Here, $\Psi_{i, \mbox{\tiny LD2}}$ are the parameter values of the LD2 scheme (e.g. $\kappa^{(4)} = 1/1024$, $\alpha = 0.36$), whereas $\Psi_{i, \mbox{\tiny Ref}}$ corresponds to standard central-scheme parameters typically used in RANS computations (e.g. $\kappa^{(4)} = 1/64$, $\alpha = 0$).
The blending function $\sigma$ is adopted from \cite{Travin2002} and discerns between the well-resolved vortex-dominated flow regions (\textit{LD2}) and coarse-grid irrotational regions (\textit{Ref}). 

By now, the hybrid LD2 scheme (HLD2) has been successfully applied in a number of hybrid RANS-LES computations ranging from canonical flows on structured grids \cite{Probst2016} to complex high-lift aircraft on mixed-element unstructured meshes \cite{Probst2022}.

\section{Basic Validation of Embedded WMLES}\label{sec:basic_validation}
Before analyzing the embedded WMLES approach from Sec.~\ref{sec:num_meth} for a complex transonic aircraft configuration with UHBR nacelle in Sec.~\ref{sec:results}, we investigate and demonstrate its basic scale-resolving functionalities in fundamental test cases, i.e. decaying isotropic turbulence for pure LES and a developing flat-plate boundary layer for WMLES.

\subsection{Decaying Isotropic Turbulence}\label{sec:dit}
Although SST-based IDDES is a well-known hybrid model present in many CFD codes, a proper verification for a given flow solver and the applied numerical scheme requires fundamental tests of the different modelling modes. 
This includes the pure LES functionality, where the hybrid model acts as Smagorinsky-type sub-grid model and mostly relies on the "outer-flow" calibration constant of SST-based IDDES, i.e. $C_{\mbox{\tiny DES}} = 0.61$.\footnote{Note that the calibration constant in SST-based DES-variants takes a different value close to walls, but this region is usually treated in RANS mode anyway..}

For this reason, we present for the first time TAU simulations of decaying isotropic turbulence (DIT) using SST-IDDES with the LD2 scheme and compare the results with classic experimental data from \cite{comte1971simple}.
In particular, the turbulent-kinetic-energy (TKE) spectra at two different time levels after the start of decay, i.e. $t = 0.87$ s and $t = 2.0$ s, are considered.
Additionally, to emphasize the effect of the LD2 scheme, further SST-IDDES simulations are performed using a reference central-scheme with higher artificial dissipation (cf. Eq. \ref{eq:num_blending} in Sec. \ref{subsec:LD2}).

As for the computational setup, a cubic domain with normalized edge length of $2\pi$ is discretized by Cartesian meshes with $32^3$, $64^3$ and $128^3$ cells, respectively.
Periodic boundary conditions are applied in all three directions.
The initial velocity field has been generated by a Kraichnan-type synthetic turbulence approach \cite{Kraichnan1970} and retains the TKE spectrum of the experiment at $t = 0$ s.
Due to the compressible formulation of the DLR-TAU code, appropriate initial density and pressure fields are derived from the isentropic relations of compressible fluids, describing the change of state from stagnation ($\mbox{Ma}_\infty = 0$) to the local Mach number, i.e. $\rho/\rho_\infty = f\left( \mbox{Ma} \right)$ and $p/p_\infty = f\left( \mbox{Ma} \right)$.
Moreover, the initial fields of modeled TKE and specific dissipation rate $\omega$ are computed in a preliminary steady-state SST-IDDES computation, where all equations except for the hybrid turbulence model are frozen.
The temporal resolutions of $\Delta t / s \in \{ \, 5\cdot10^{-3}, \, 5\cdot10^{-3}, \,  2\cdot10^{-3} \}$ for the coarse, middle and fine grid were determined in time-step convergence studies.

\begin{figure}
	\vspace{0cm}
	\begin{center}
	\begin{minipage}[c]{0.48\textwidth}
			\includegraphics[trim= 130 50 200 200, clip,width=\linewidth]{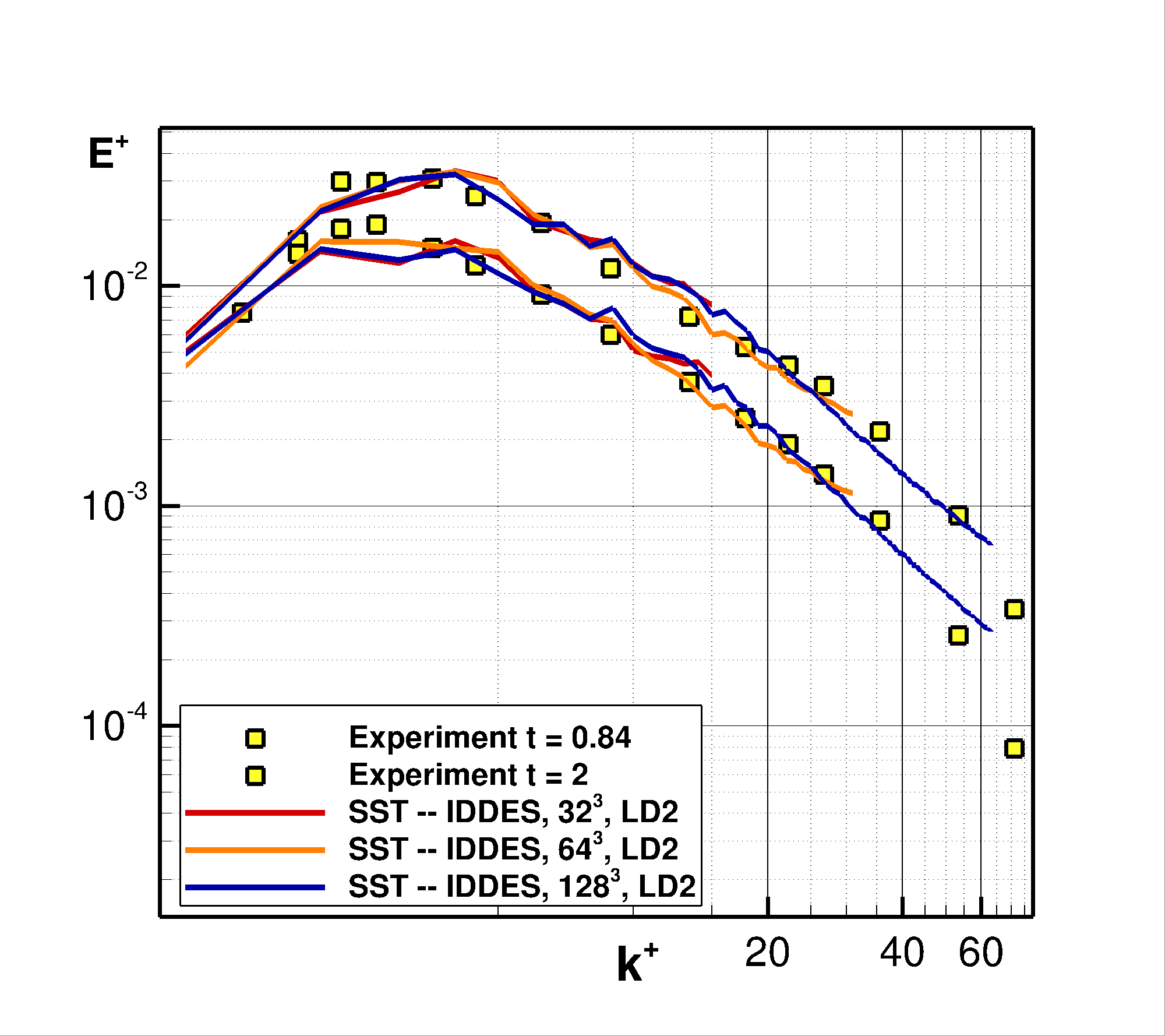}	
	\end{minipage}%
	\hspace*{0.3cm}
	\begin{minipage}[c]{0.48\textwidth}
			\includegraphics[trim= 130 50 200 200, clip,width=\linewidth]{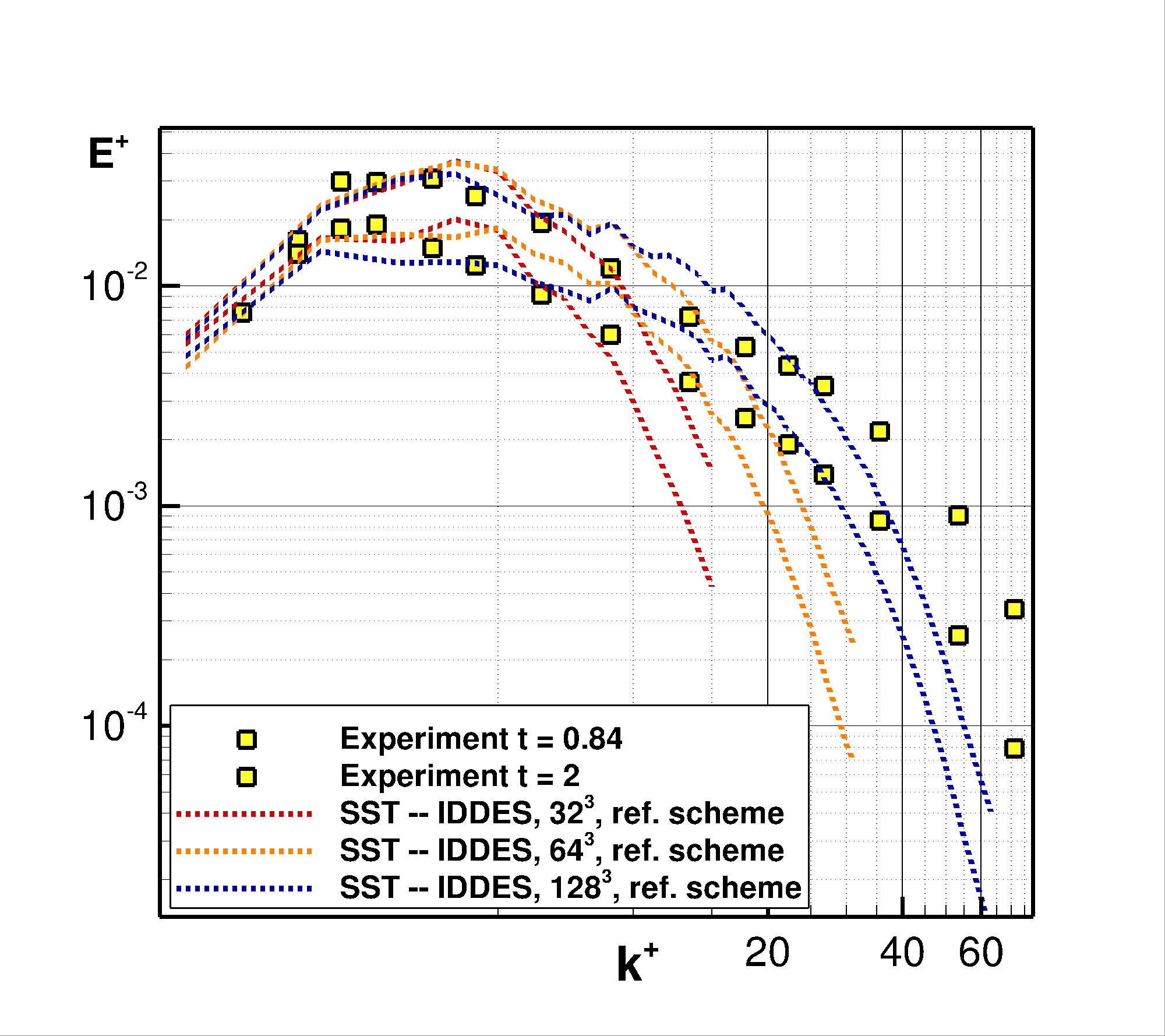}	
	\end{minipage}%

	\caption{TKE spectra of decaying isotropic turbulence (DIT) for two different times along with experimental data \cite{comte1971simple}. Results for the LD2 scheme (left) and a reference central-scheme (right) are shown.}
	\label{abb:DIT}
	\end{center}
\end{figure}

Fig. \ref{abb:DIT} (left) shows the results for the SST-IDDES with LD2 scheme which demonstrate a good agreement with the experimental results for all spatial resolutions and both time levels. For the reference central-scheme however, the picture is different. 
Although there are agreements with the experimental results for small wave numbers scales $k^+ \leq 8$ for all resolutions and time levels, deviations arise for larger wave numbers.
These deviations are growing with increasing wave number and finally result in a significant underestimation of the TKE for all setups.

As a result we successfully demonstrated the LES functionality of SST-IDDES in conjunction with the LD2 scheme. The low dissipation feature of the numerical scheme was confirmed and additionally emphasized by reference simulations with higher artificial dissipation.

\subsection{Developing Flat Plate Boundary Layer}
\label{sec:flat_plate}
For a basic assessment of the full embedded WMLES functionality, we consider the test case of a developing flat-plate boundary layer, which transitions from RANS to WMLES at a fixed streamwise position.
It starts with zero thickness at the inflow and is computed in SST-RANS mode up to 
the position, where the momentum-thickness Reynolds number reaches $Re_\theta = 3040$.
Here, a zonal switch to WMLES within IDDES is placed, along with a synthetic-turbulence forcing region of about half a boundary layer thickness in streamwise direction, see Sec.~\ref{subsec:stg}.

A hybrid grid with 5.8 million points and hexahedral cells in the WMLES area is used, which ensures $\Delta x^+ \approx 100-200$, $\Delta y^+ \approx 1$, $\Delta z^+ \approx 50$ like the structured grid used in \cite{Probst2018}. 
More relevant for WMLES, the streamwise spacing fulfills $\Delta x  \leq  \delta / 10 $ throughout the flow domain, where $\delta$ is the approximate local boundary layer thickness.
The normalized timestep (in wall units) is $\Delta t^+ \approx 0.4$ and safely fulfills the convective CFL criterion ($\mbox{CFL}_{conv} \lt 1$) in the whole LES region.

The statistical input data for the STG methods is given by external input from a precursor RANS profile at $Re_\theta = 3040$ which has been augmented with an anisotropic normal-stress approximation according to
\cite{Laraufie2013}.

\begin{figure}
	\vspace{0cm}
    \begin{center}
	\begin{minipage}[c]{0.5\textwidth}
			\includegraphics[trim= 80 80 140 140, clip,width=\linewidth]{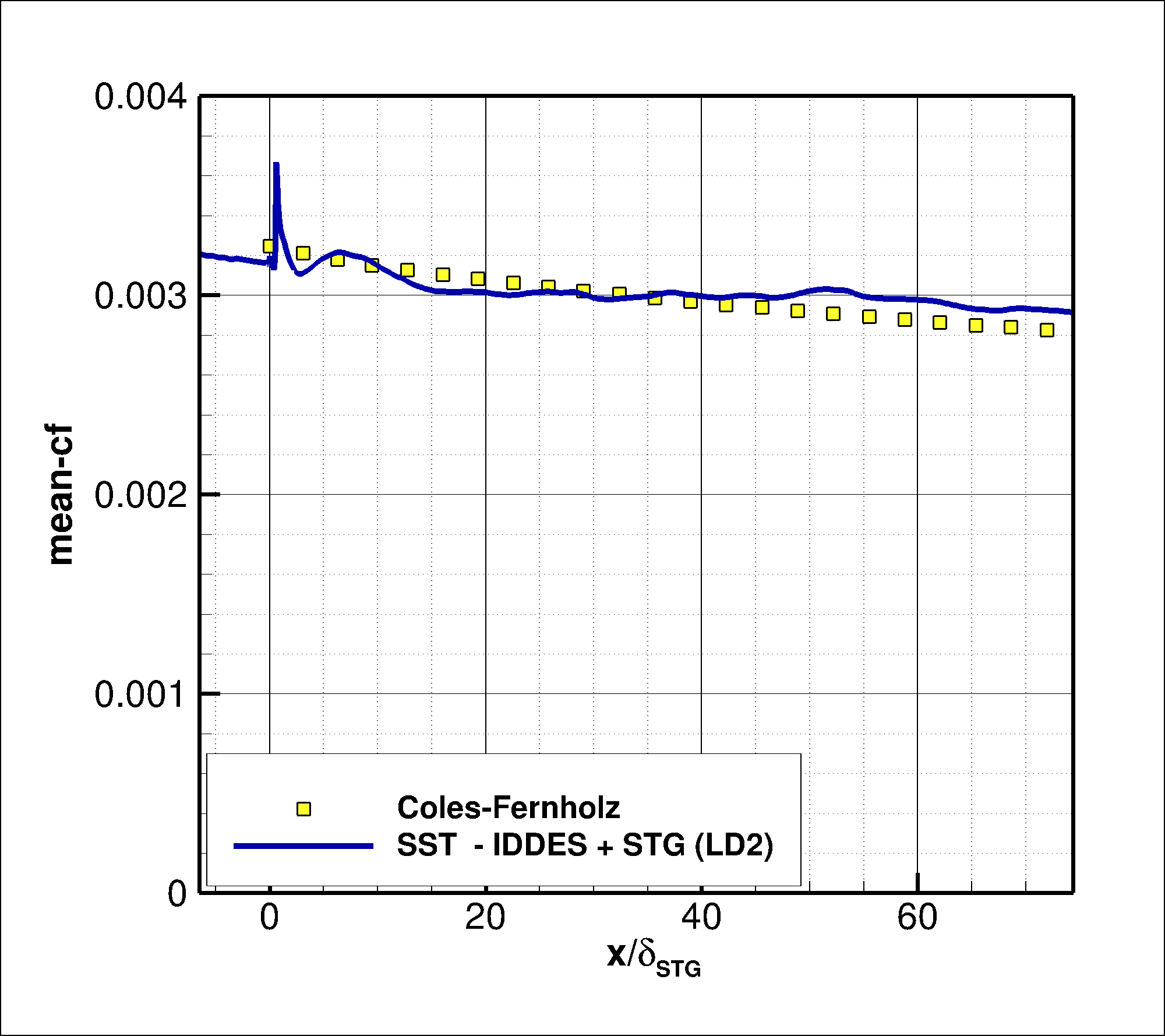}	
	\end{minipage}%

	\caption{Evolution of averaged skin friction along streamwise position $x$ of the flat plate test case.}
	\label{abb:flat_plate}
	\end{center}
\end{figure}

The spanwise and temporal averaged results of the skin friction distribution mean-$c_f$ are depicted in Fig. \ref{abb:flat_plate} along with the Coles-Fernholz correlation \cite{nagib2007approach}.
After an initial overshoot of mean-$c_f$ at the position of the STG, mean-$c_f$ shows good agreement with the Coles-Fernholz correlation and remains within an acceptable error margin of $5\,\%$.
Note that the adaption region downstream of the STG is hardly visible but still present. This region is defined as underprediction of mean-$c_f$ compared to the previous mean-$c_f$ level directly upstream of the STG. The adaption-length which respresents the distance between the position of the STG and the first peak in mean-$c_f$ downstream of the overshoot amounts $7\,\delta_{STG}$ where $\delta_{STG}$ is the boundary layer thickness at the position of the STG.
Within this adaption region the sum of modelled and resolved turbulent stresses are lower than the previous level of modelled turbulence of the RANS region which results in an underprediction of mean-$cf$ \cite{franccois2020development}.

Finally, this examination confirms the embedded WMLES functionality of SST-IDDES with STG for a flat plate flow. Thus this methodic is basically verified for comparable geometry sections at the XRF-1-UHBR configuration.

\section{Grey-Area Investigation on Nacelle-Aircraft Configuration}
\label{sec:results}

\subsection{Geometry, Flow Conditions and RANS Mesh}
\label{sec:geometry}
\begin{figure}
	\vspace{0cm}
	\hspace{3cm}
	\begin{minipage}[c]{0.48\textwidth}
			\includegraphics[clip,width=\linewidth]{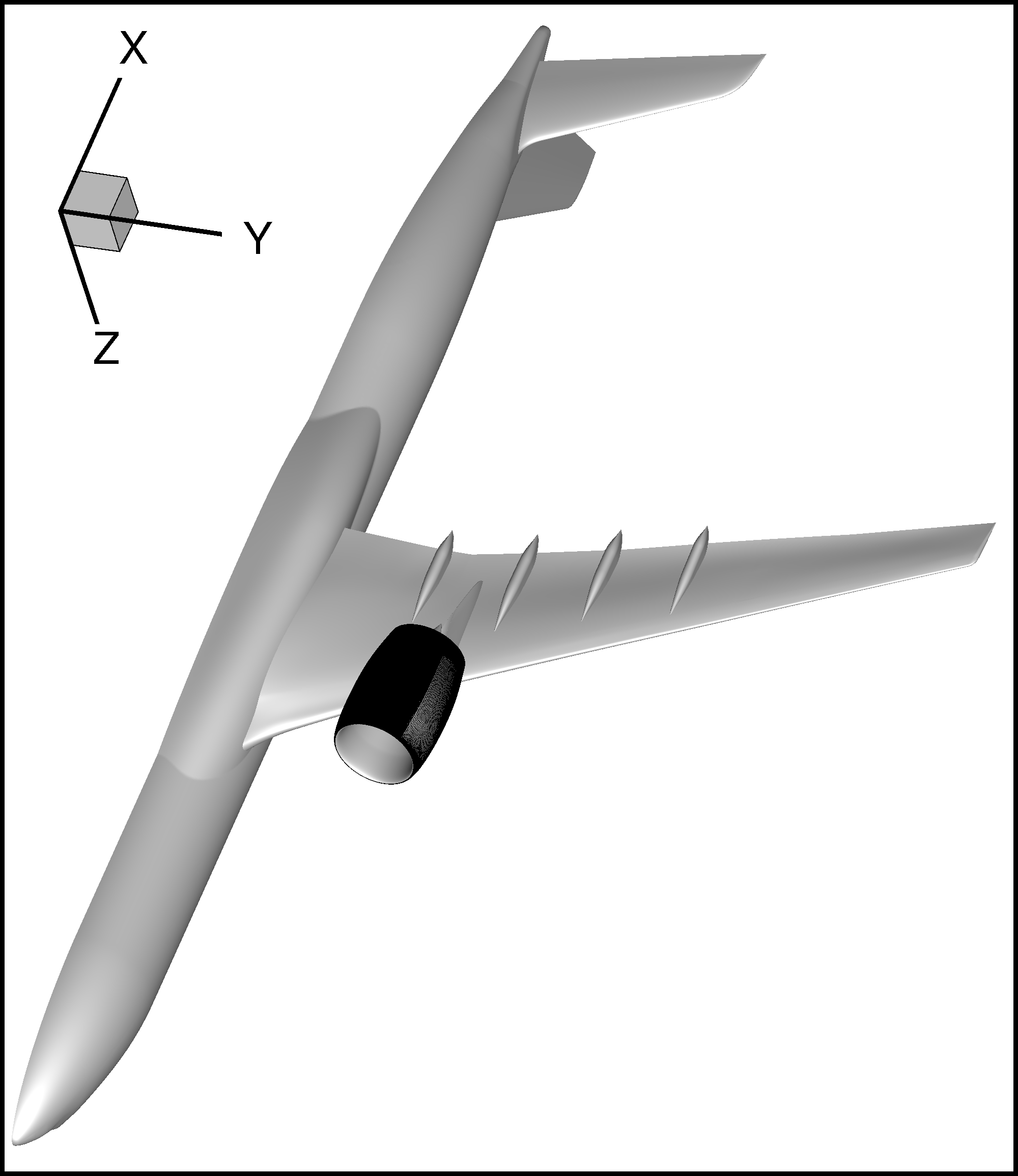}	
	\end{minipage}%

	\caption{Bottom view of XRF-1 - aircraft configuration with UHBR nacelle. The nacelle lower side includes the mesh refinement region for embedded WMLES.}
	\label{abb:XRF1_overview}
\end{figure}

The actual target configuration consists of a half model of a modern transport aircraft configuration in conjunction with a through flow nacelle (cf. Fig. \ref{abb:XRF1_overview}). The employed XRF-1 aircraft model represents a wide-body long-range research configuration and is designed by Airbus.
A Ultra High Bypass Ratio (UHBR) nacelle is integrated with the aid of a pylon and positioned close to the wing lower side. The UHBR design consists of an outer casing and a core body with plug. The casing is shaped circularly with a cross section similar to an airfoil. Both, nacelle and a specifically designed pylon were developed by DLR \cite{spinner2021design}. 

In order to find a suitable flow condition with shock induced separation in the surrounding of the nacelle surface a comprehensive numerical study was performed where various high speed off-design conditions were assessed. As key parameter for the occurrence of transonic shocks at a Reynolds number of $Re=3.3$ million a low angle of attack ($\alpha$) was identified. For a farfield Mach number of 0.84 and $\alpha=-4^\circ$ shock induced separation is present at the wing lower side, the pylon and the nacelle. A single, locally separated transonic shock could be found at the outer surface of the nacelle lower side (cf. Fig. \ref{abb:nacelle_RANS}). Thus, a flow condition which allows to examine an isolated shock with subsequent boundary layer separation in the context of a nacelle-aircraft configuration was found.

In a prelinimary work a high quality RANS mesh for the XRF-1 - UHBR half model was designed and constructed by projects partners of the research unit at the University of Stuttgart and DLR.
The surface RANS mesh mainly consists of structured areas which are extruded to hexahedral blocks. These are designed to contain the entire RANS boundary layer with a safety factor of 2. 
The wall adjacent cell spacing fulfills $y^+(1)\leq 0.4$ and a growth rate of 1.12 is applied in wall normal direction.
A h-type mesh topology is employed at the intersections of the aircraft components to be able to accurately resolve flow features in these areas.
The farfield region is discreticed by tetrahedra and extends to 50 wingspans in all coordinate directions. The total grid size before refinement amounts 112 million points.

\subsection{Grid Design for Embedded WMLES}
\label{sec:grid_design}
In the following the mesh design for the WMLES refinement region is introduced. 
A sophisticated meshing strategy, that aims to reduce the grid size as far as possible but follows basic refinement and extension constraints for WMLES, is developed. This is necessary in order to limit mesh size and resulting computing time to a reasonable level. 
Special care was taken to the mesh resolution of all coordinate directions ($\Delta x, \Delta y$ and $\Delta z$) which depend on the local boundary layer thickness $\delta$. Additionally, a potential shock movement is considered with regard to the refinement extension as well as mesh resolution.

The refinement region is embedded within the previously described RANS mesh with the aid of unstructured bands in the surface mesh (cf. Fig. \ref{abb:nacelle_RANS} and Fig. \ref{abb:mesh_surface_mesh}). This strategy allows to drastically increase the resolution within the structured boundary layer such that the surrounding  RANS region remains unchanged.
An unstructured nearfield block, which is also present in the pure RANS mesh, serves as an interface between the hexahedral blocks and the farfield, exhibits a mesh decay rate of $0.85$. The total mesh size of the combination of RANS mesh and refinement region for WMLES comprises 420 million points.

\subsubsection{Extension of the refinement region}
To describe locations on the nacelle surface more precisely a cylindrical coordinate system $r, \varphi \text{ and } x/c $ is introduced, where $c$ represents the nacelle chord length.
Its reference point $r=0, \ x/c = 0$ is located in the nacelle center within a cross section that includes the entire nacelle leading edge.
$\varphi$ is set to $0^\circ$ at the intersection between nacelle and pylon and increases in clockwise direction that $90^\circ$ points towards the fuselage.

\begin{figure}
	\vspace{0cm}
	\begin{minipage}[c]{0.48\textwidth}
			\includegraphics[clip,width=\linewidth]{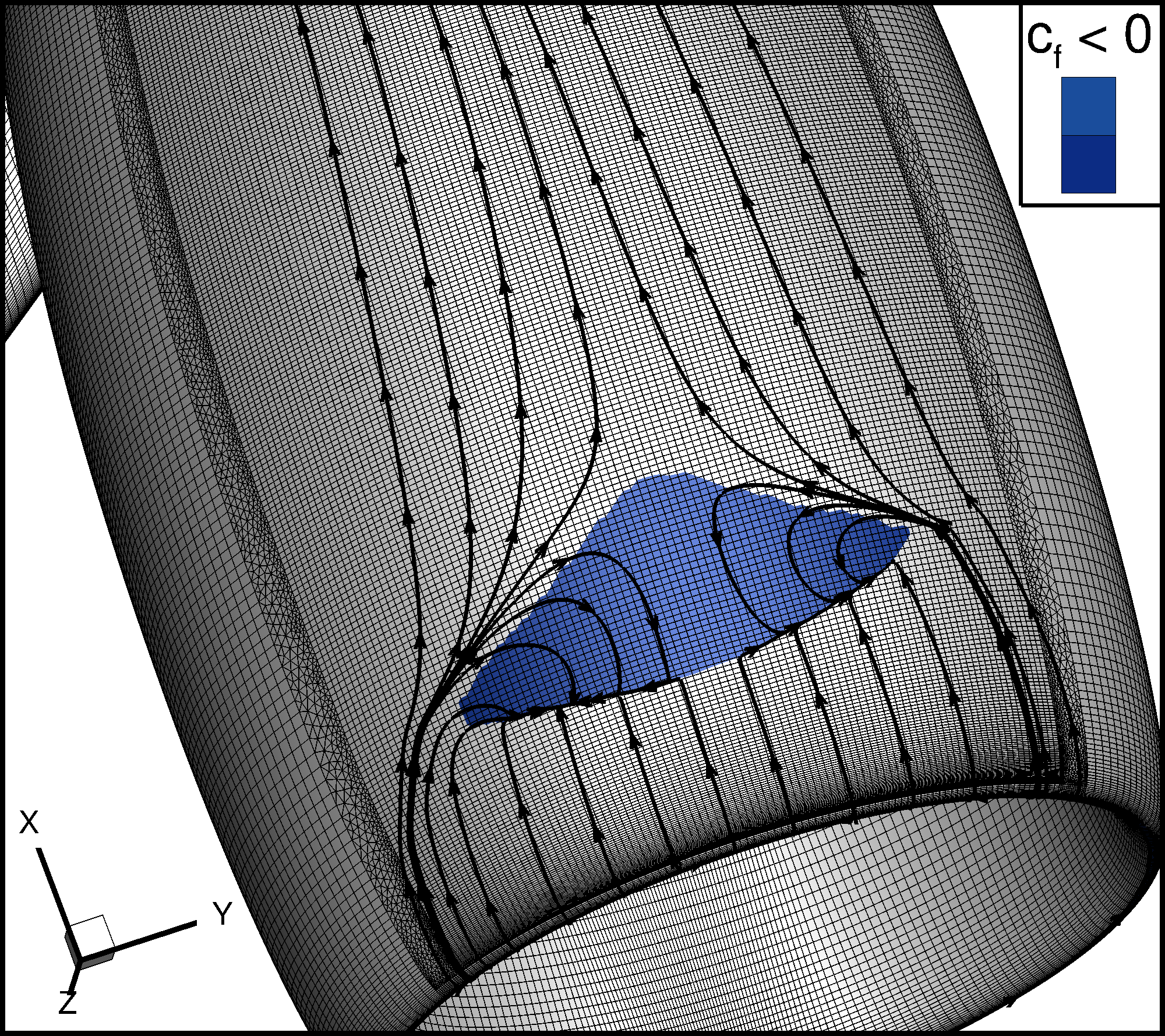}	

	\end{minipage}%
	\hspace*{0.2cm}
	\begin{minipage}[c]{0.5\textwidth}
		\includegraphics[clip,width=\linewidth]{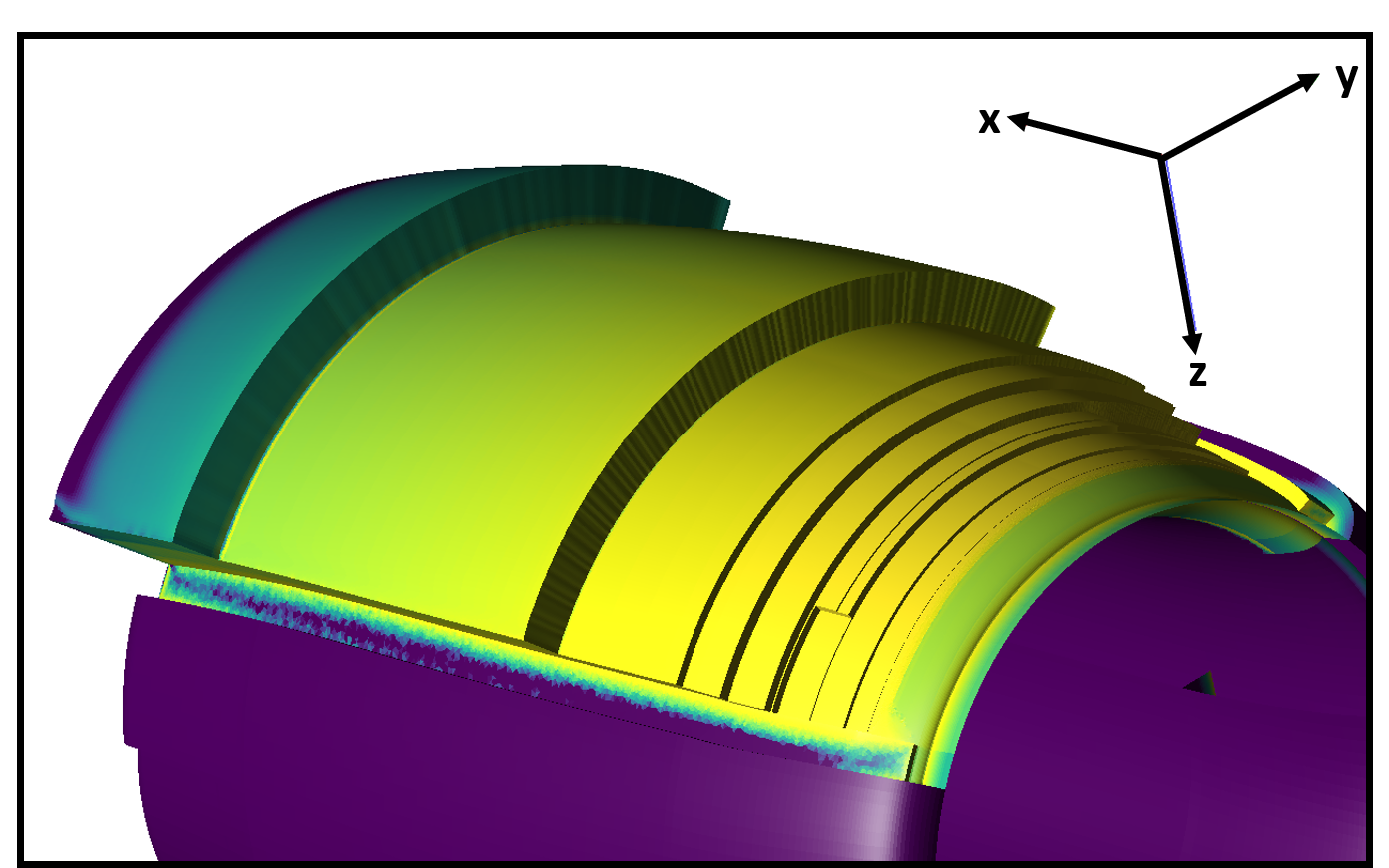}%
	\end{minipage}%
	\caption{ Bottom view of the UHBR-nacelle. \textbf{Left:} Area of recirculation of SST-RANS solution for $\mathbf{Ma_{\infty}=0.84}$ and $\mathbf{\alpha=-4^\circ}$. The shown RANS surface mesh already includes the boundaries for the refinement region in form of unstructured streaks. \textbf{Right:} Extension of refinement area with stepwise increase in streamwise direction. The colorbar visualizes the cell surface area where yellow and purple represent large and low areas, respectively.}
	\label{abb:nacelle_RANS}
\end{figure}

According to \citep{spalart2001young} the first step in designing hybrid RANS LES mesh for DES based algorithms is the definition of the RANS and LES regions for the given configuration.
Since the aim of this research topic is the application of a WMLES methodology to a flow region with shock induced separation, all flow regions directly related to this phenomenon are of interest and should be highly resolved.
The primary region is the area of recirculation (AOR) downstream of the shock position (cf. Fig. \ref{abb:nacelle_RANS} left). Flow regions related to this are the attached boundary layer upstream of the AOR and separated boundary layer downstream of the AOR until the trailing edge of the nacelle.
To this end the average shock front position and extension of the AOR are calculated by a preceding SST-RANS calculation. Fig. \ref{abb:nacelle_RANS} (left) shows a surface plot of the skin friction coefficient ($c_f$) where the $c_f$ is only plotted for $c_f<0$ which serves as an indicator of the AOR.
 The refinement region in spanwise direction  ($\varphi$) is chosen such that the entire area of recirculation is included with some margins in $\varphi$-direction and extends $105^\circ$ starting from  $120^\circ$ until $225^\circ$ (cf. Fig. \ref{abb:nacelle_RANS}).

Since the boundary layers thickness is not only a function of $x$ but also of $\varphi$ we introduce the new variables $\delta_{\varphi, max}(x)$ and $\delta_{\varphi, min}(x)$ which refer to the maximum and minimum boundary layer thickness for a given streamwise position $x$.
In $x/c$ direction the refinement is applied between $x_a/c=0.06$ and $x_b/c=1$.
The choice of $x_a/c=0.06$ as the most upstream position is the result of the dependence of mesh resolution on the boundary layer thickness $\delta_{\varphi, min}(x)$. 
The smaller the boundary layer thickness $\delta_{\varphi, min}(x)$ at location $x_a$ the smaller the required cell lengths $\Delta \zeta(x_a)$ for $\zeta \in \{r,\varphi,x\}$ since $\Delta \zeta(x) \leq \delta_{\varphi, min}(x)/10$. 
The refinement in wall normal direction $r$ is applied for wall distances that hold $ d_w(x) \leq 1.2\cdot \delta_{\varphi, max}(x)$ in the interval $0.06 \leq x/c \leq 0.16$ and $d_w \leq 1.5\cdot \delta_{\varphi, max}(x)$ within $0.16 \leq x/c \leq 1$. Thus $d_w/c$ ranges from  $0.2\%$ at $x/c=0.06$ to $15\%$ at the trailing edge (cf. Fig. \ref{abb:nacelle_RANS} right). 
Although these distances are smaller than  $ d_w \leq 2\cdot \delta(x)$ suggested by \cite{menter2012best} we show in Sec. \ref{sec:transient_process} that the whole resolved boundary layer remains within the refined area with distance $d_{refined}(x)$ over the entire simulated time period. Additionally, the extension of the refinement area in $r$-direction also considers a potential oscillation of the boundary layer separation point around its average position at $x_s/c=0.13$ (SST-RANS solution). We assumed an oscillation amplitude of $\pm 0.03\,c$ which also allows to employ this mesh in case of shock buffet. 
As a consequence, at position $x/c=0.16$ a refinement distance of ${d_{refined}(0.16c) =  1.2 \cdot \delta_{\varphi, max}(0.19c)}$ is used.

\subsubsection{Resolution of the refinement region}
\label{sec:mesh_resolution}

\begin{figure}
	\vspace{0cm}
	\begin{minipage}[c]{0.48\textwidth}
			\includegraphics[clip,width=\linewidth]{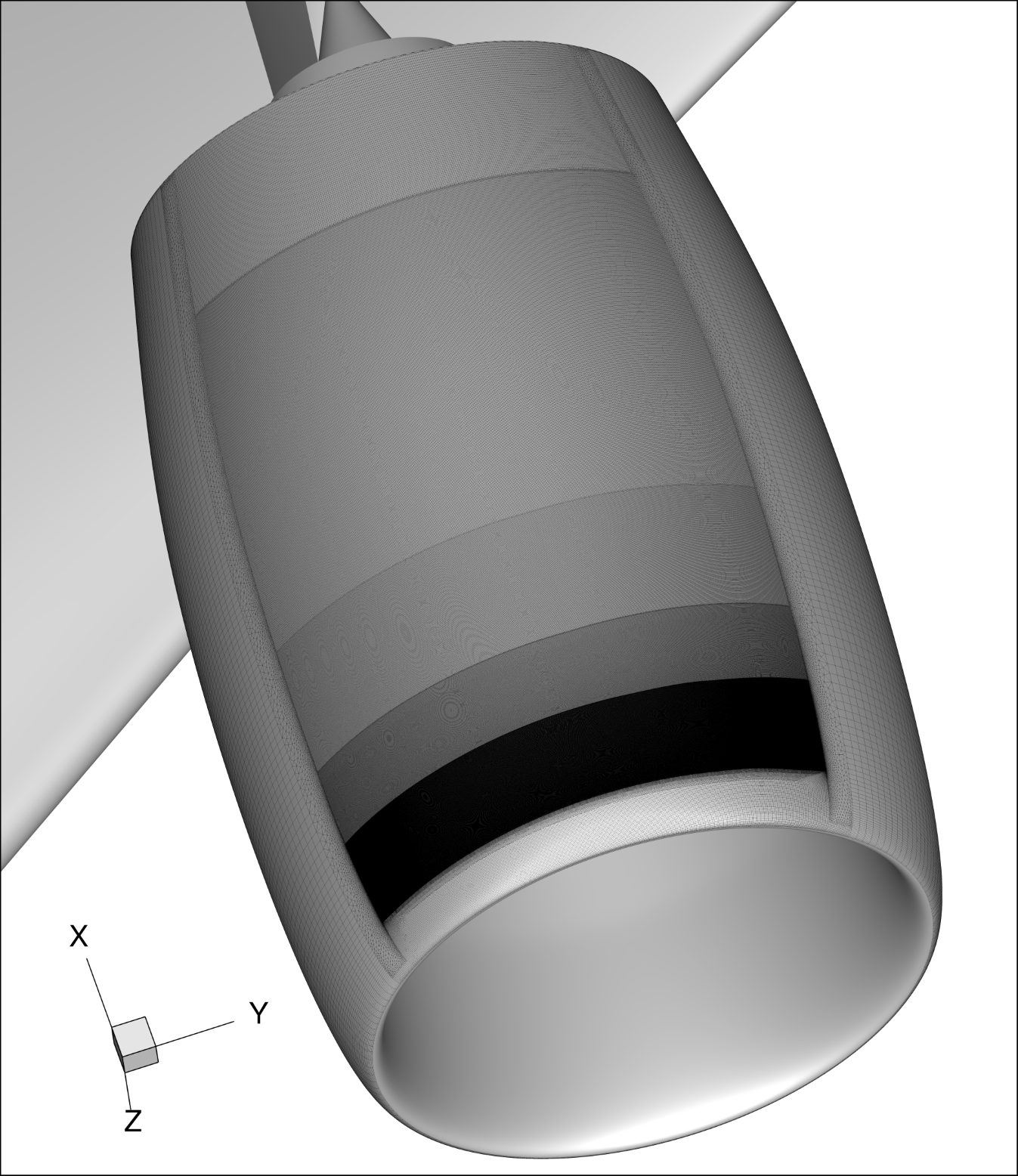}	

	\end{minipage}%
	\hspace*{0.2cm}
	\begin{minipage}[c]{0.48\textwidth}
		\includegraphics[clip,width=\linewidth]{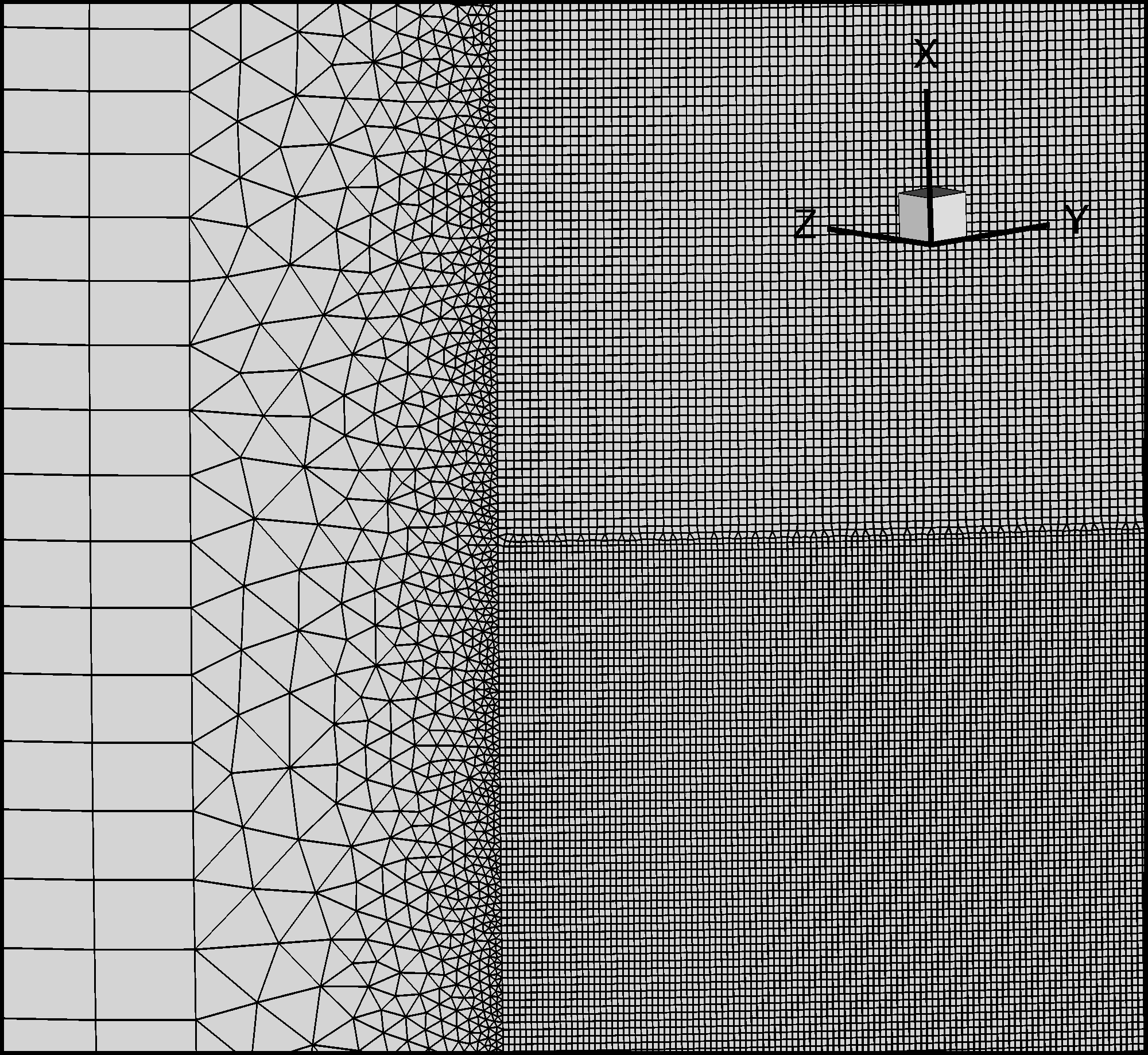}%
	\end{minipage}%
	\caption{Surface mesh of refinement region on lower side of UHBR nacelle. \textbf{Left: } Discrete coarsening of $\Delta \varphi$ is apparent which subdivides the refinement area into five subregions. \textbf{Right: } Vertical unstructured (triangular based) streak enables to refine locally and keep surrounding RANS resolution untouched. Horizontal unstructured stripe allows to coarsen the refinement region in $\varphi$-direction.}
	\label{abb:mesh_surface_mesh}
\end{figure}

\begin{figure}
	\vspace{0cm}
	\begin{minipage}[c]{1\textwidth}
			\includegraphics[trim= 180 130 310 330,clip,width=\linewidth]{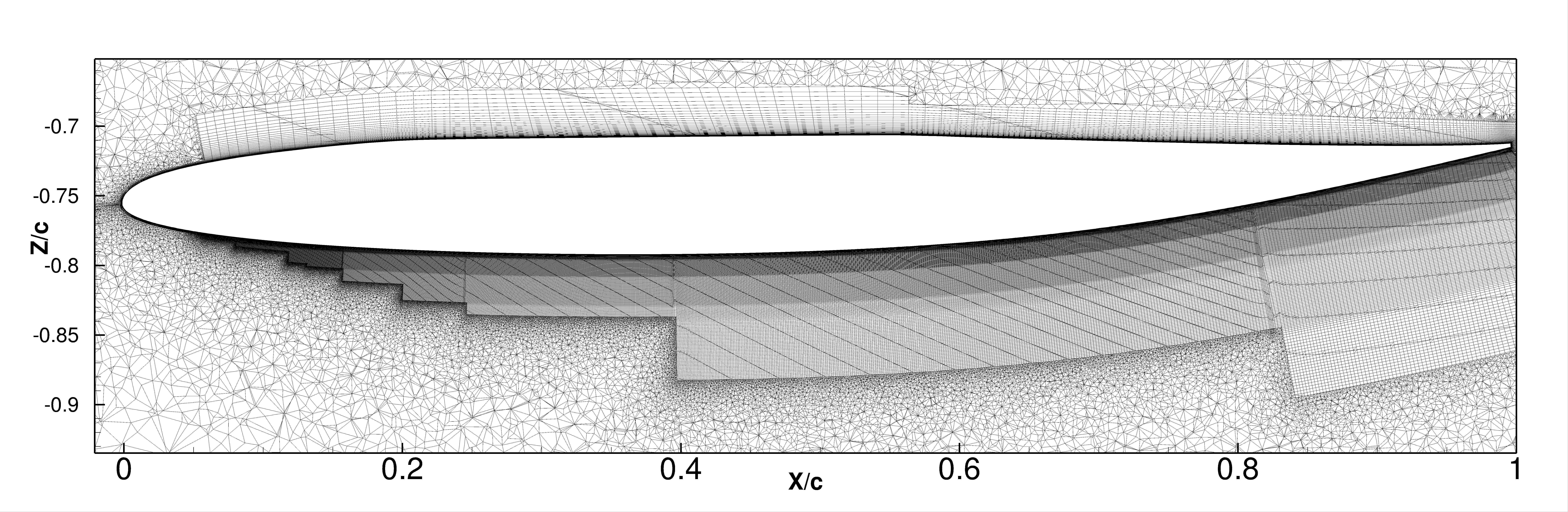}	
	\end{minipage}%

	\caption{Cross section of nacelle lower side at $\varphi= 180^\circ $. Subregion $\Omega_1$ ($ 0.06 \leq x/c \leq 0.16$) of the refinement region includes 200 Mio. cells which corresponds to $48\%$ of the entire grid size.}
	\label{abb:mesh_cross_section}
\end{figure}

The resolution in $x$-direction depends on the local boundary layer thickness and is set to a limit of $\Delta x (x) \leq \delta_{\varphi, min}(x) / 10 $ which leads to a total number of 1350 points in $x$-direction from the leading edge to the trailing edge. Again an oscillation of separation due to shock buffet point is considered. Thus it is assumed to have a attached boundary layer until $x_s/c=0.13+0.03$ leading to reduced boundary layer thickness compared to the preliminary SST-RANS solution. Therefore the boundary layer thickness at $x/c=0.16$ is estimated to ${\delta_{\varphi, min}(x/c=0.08) \cdot   2^{4/5}}$ according to turbulent boundary layer theory.
As before the resolution in $\varphi$-direction is limited to $r \Delta \varphi(x) \leq \delta_{\varphi, min}(x) / 10$. In contrast to the resolution in $x$-direction the adaption of $\Delta \varphi(x)$ to $\delta_{\varphi, min}(x)$ is realised in a discrete manner. Therefore the refinement region is separated into five subregions with its boundaries located at $x/c \in \{0.06; 0.16; 0.25; 0.4; 0.82; 1\}$ (cf. Fig. \ref{abb:mesh_surface_mesh}).  $\Delta \varphi(x)$ remains constant within each subregion $\Omega_i$ and is set to ${r \Delta \varphi(x \in \Omega_i)=\delta_{\varphi, min}(x_i) / 10 }$ with $x_i$ defined as the most upstream position of $\Omega_i$. With this protocol the resolution in $\varphi$-direction is always smaller than $\delta_{\varphi, min}(x) / 10 $ which results into $\{4350; 1660; 870; 603; 250\}$ points in $\varphi$-direction within the corresponding subregion. Without this stepwise increase of $\Delta \varphi$ the total grid number would increase by a factor of 3 to $1.2\cdot 10^9$ points.
Again a potential movement of the boundary layer separation point is considered and therefore ${r\Delta \varphi (x=0.16c) =\frac{1}{10} \delta_{\varphi, min}(x=0.08c) \cdot  2 ^{4/5}}$.
In $r$-direction the wall normal spacing of the wall adjacent cells is limited to $r^+(1)=0.4$. The cells of the entire refinement area are extruded geometrically with a growth factor of 1.12 until $\Delta r=\Delta x(x=0.06c)$ is reached and $\Delta r$ is initially kept constant to obtain locally isotropic cells. Since the distance of the refinement region $d_{refined}(x)$ increases in $x$-direction in a cascading manner (cf. Fig. \ref{abb:nacelle_RANS} (right) and \ref{abb:mesh_cross_section}) the geometric growth is continued for refinement areas with larger wall distances.
Exemplarily, $\Delta r$ is further increased to $\Delta r=\Delta x(x=0.16c)$ for wall distances in the interval $ d_{refined}(x=0.16c) \leq r \leq d_{refined}(x=0.25c)$ and applied where $0.16 \leq x/c \leq 1$. Subsequently $\Delta r$ is again increased until $\Delta r=\Delta x(x=0.25c)$ for wall distances in the intervall $ d_{refined}(x=0.25c) \leq r \leq d_{refined}(x=0.4c)$ and applied where $0.25 \leq x/c \leq 1$. This protocol is repeated until $\Delta r$ amounts $\Delta r=\Delta x(x=0.82c)$ for $ d_{refined}(x=0.82c) \leq r \leq d_{refined}(x=1c)$ and $ 0.82 \leq x/c \leq 1$. Finally, the total number of grid points in wall normal direction  comprises $\{113; 168; 183; 230; 258\}$ points within the corresponding subregion.

\subsection{Results of Transient WMLES Establishment}
\label{sec:transient_process}
 As initial solution for the SST-IDDES a converged SST-RANS solution was employed. The physical time step size amounts $\Delta t = 5.5\cdot 10^{-8}\,\text{s}= 1/16750\,\text{CTU}$ where $1\,\text{CTU}=u_\infty \cdot c$ represents a single convective time unit (CTU). $\Delta t$ is chosen that $\text{CFL}<1$ is fulfilled for all grid cells.

\begin{figure}
	\vspace{0cm}
	\begin{minipage}[c]{1\textwidth}
			\includegraphics[trim= 15 5 5 30,clip,width=\linewidth]{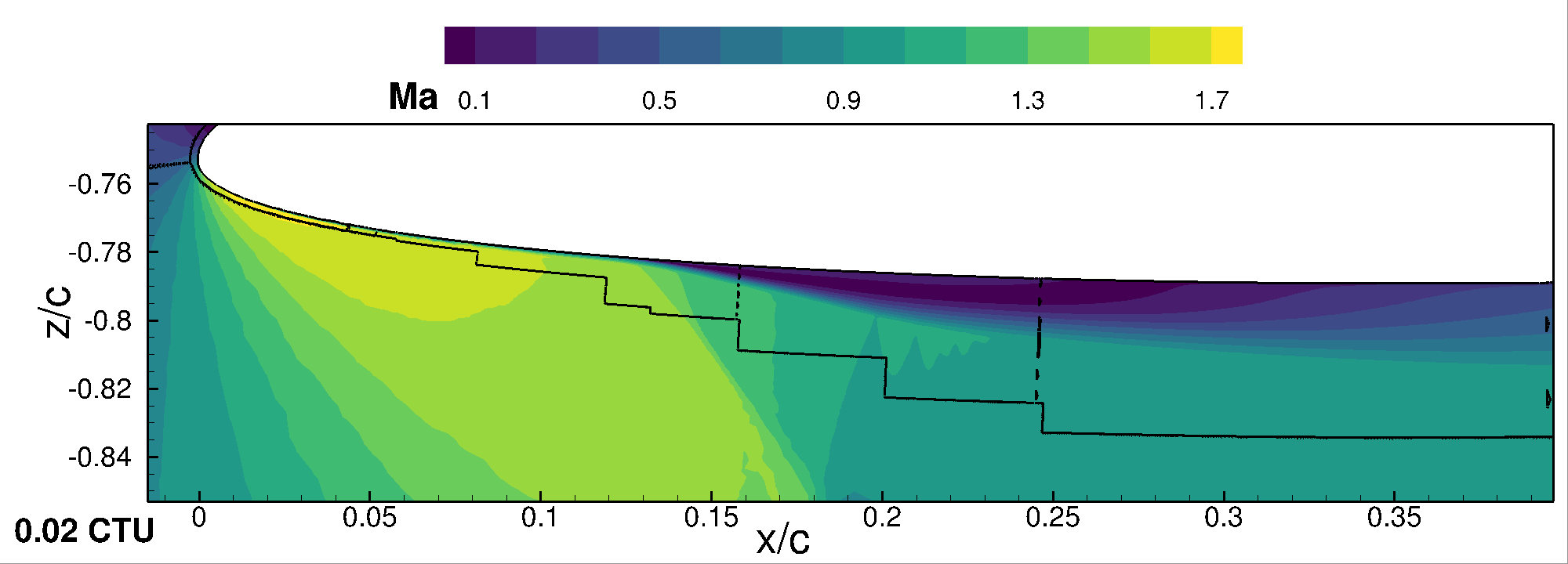}	
	\end{minipage}%
	
	\begin{minipage}[c]{1\textwidth}
			\includegraphics[trim=15 5 5 120,clip,width=\linewidth]{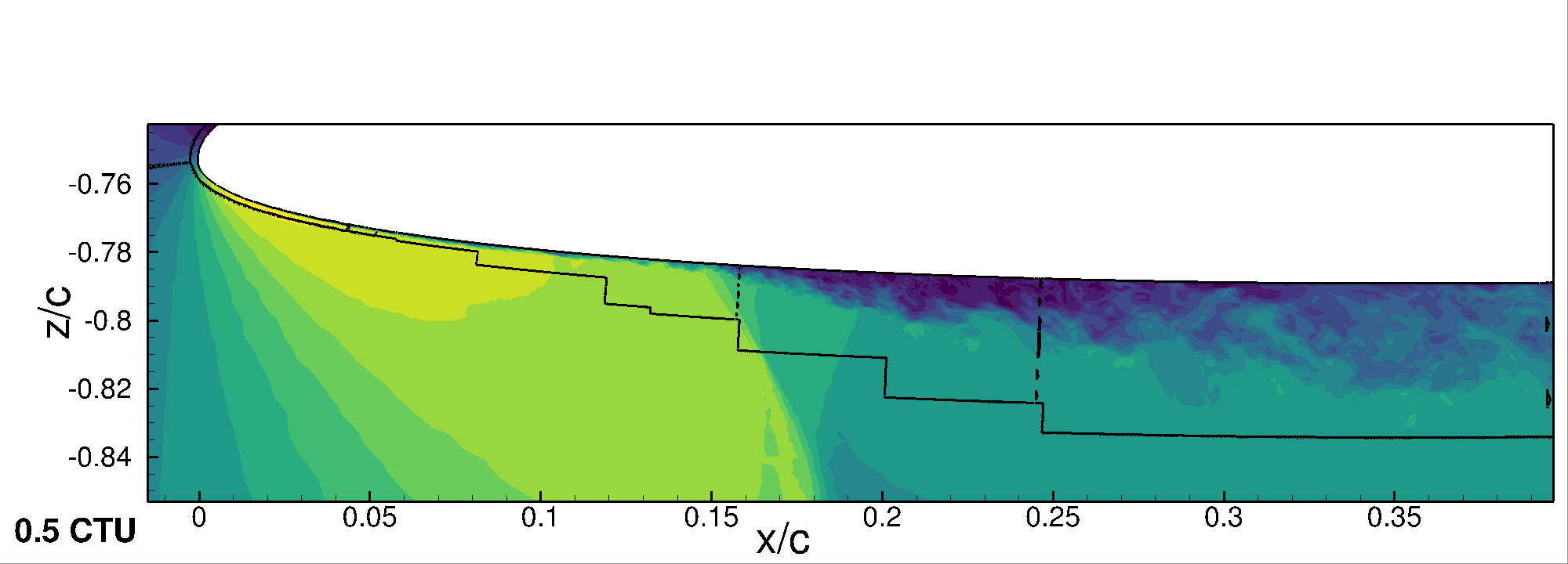}	
	\end{minipage}%
    
    \begin{minipage}[c]{1\textwidth}
			\includegraphics[trim=15 5 5 120,clip,width=\linewidth]{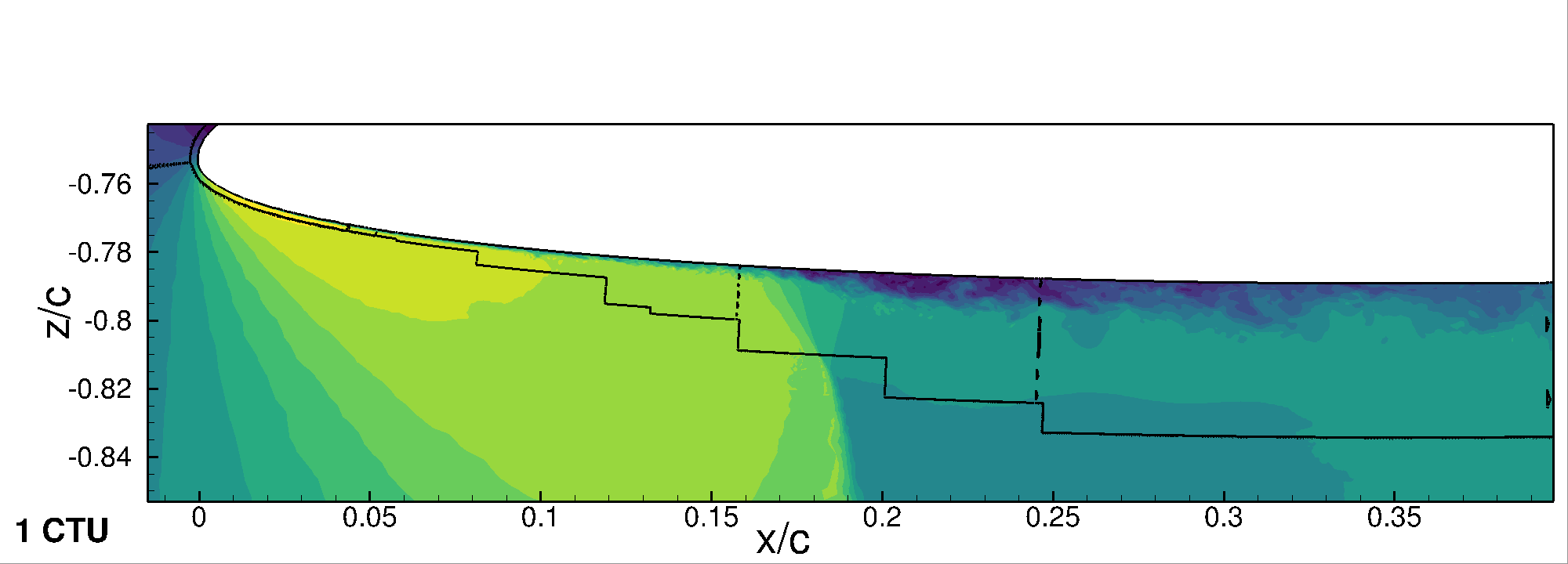}	
	\end{minipage}%
	
	\begin{minipage}[c]{1\textwidth}
			\includegraphics[trim= 15 5 5 120,clip,width=\linewidth]{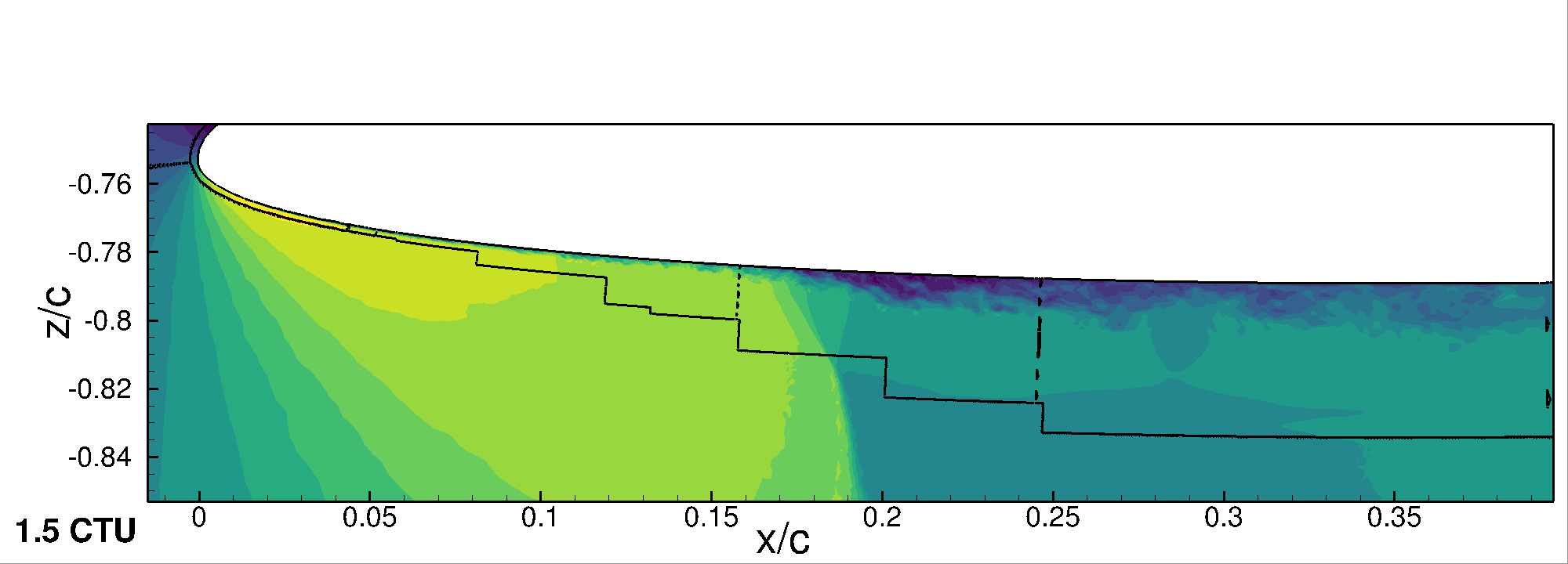}	
	\end{minipage}%
	\caption{ $Ma$-number fields within a cross section of the refinement volume at $\varphi=180^\circ$ for four different times. }
	\label{abb:snapshot_mach}
\end{figure}

Fig. \ref{abb:snapshot_mach} represents the temporal evolution of the Mach number
in a cross section at $\varphi=180^\circ$ and four different times. 
With regard to the turbulent boundary layer thickness $\delta$ it should be noted that $\delta$ is entirely located within the refinement volume with sufficient distance to its boundary (indicated by black lines). After the depicted maximal extension at $0.5\,$CTU the boundary layer thickness significantly decreases at later times. This decrease appears to be related with the shock movement in downstream direction since this correlation is also observed for various transonic flows of wing profiles \cite{jacquin2009experimental}. As mentioned before the root of the shock front $x_s$ is moving from its initial SST-RANS position $x_s(t_0)=0.13c$ downstream to $x_s(t_{1\,\text{CTU}})=0.17c$ and remains at the same position until $x_s(t_{1.5\,\text{CTU}})$. Although $x_s$ is located further downstream as we assumed for the mesh design ($0.1 \leq x_s/c \leq 0.16$) one has to note that such shock displacements are common in transient simulations (e.g. $t\leq 7.5\,\text{CTU}$). The shock position will most likely move upstream again for more advanced simulation times. %

\begin{figure}
	\vspace{0cm}
	\begin{center}
	\begin{minipage}[c]{0.6\textwidth}
			\includegraphics[trim= 70 5 60 30,clip,width=\linewidth]{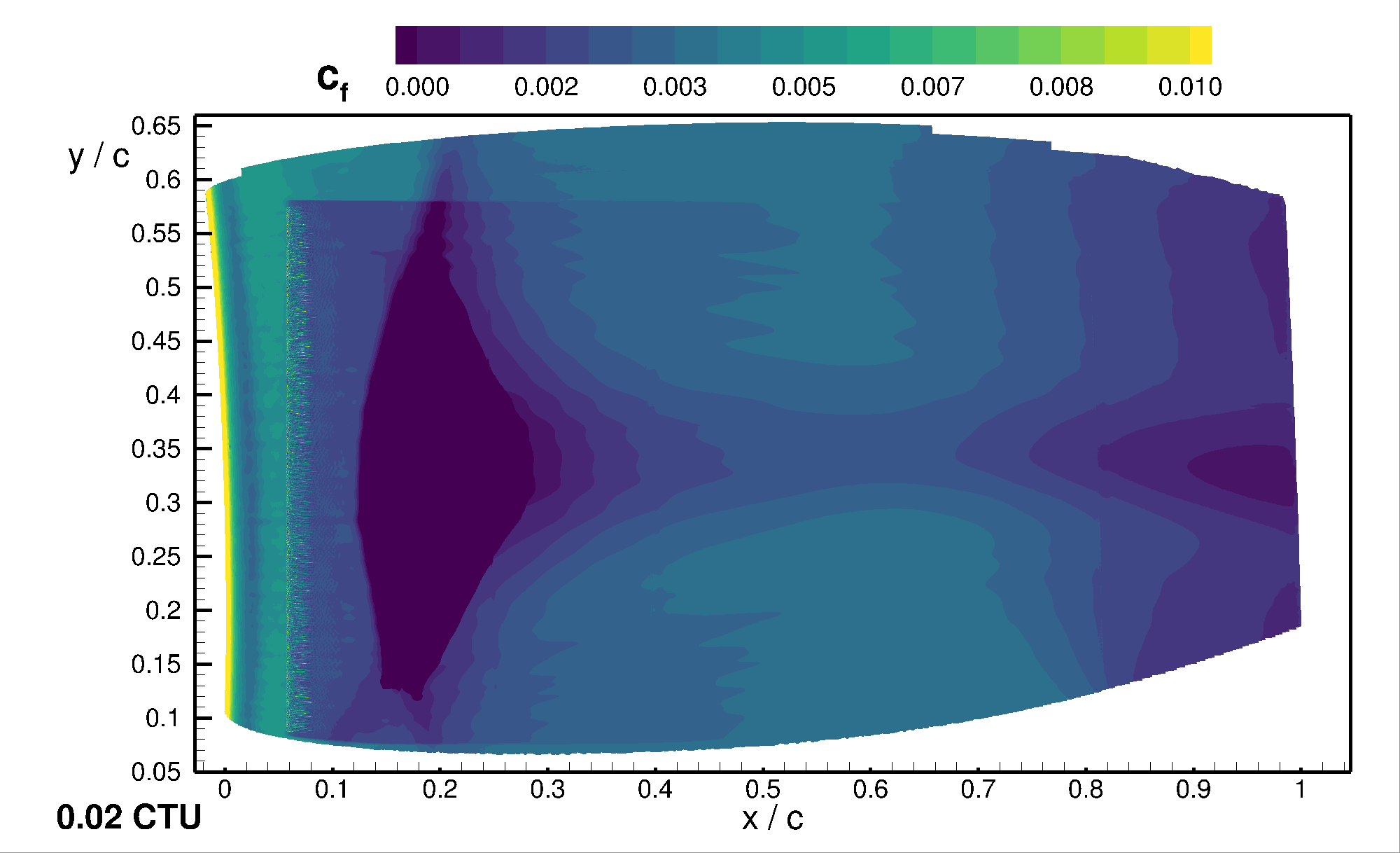}	
	\end{minipage}%
	
	\begin{minipage}[c]{0.6\textwidth}
			\includegraphics[trim= 70 5 60 160,clip,width=\linewidth]{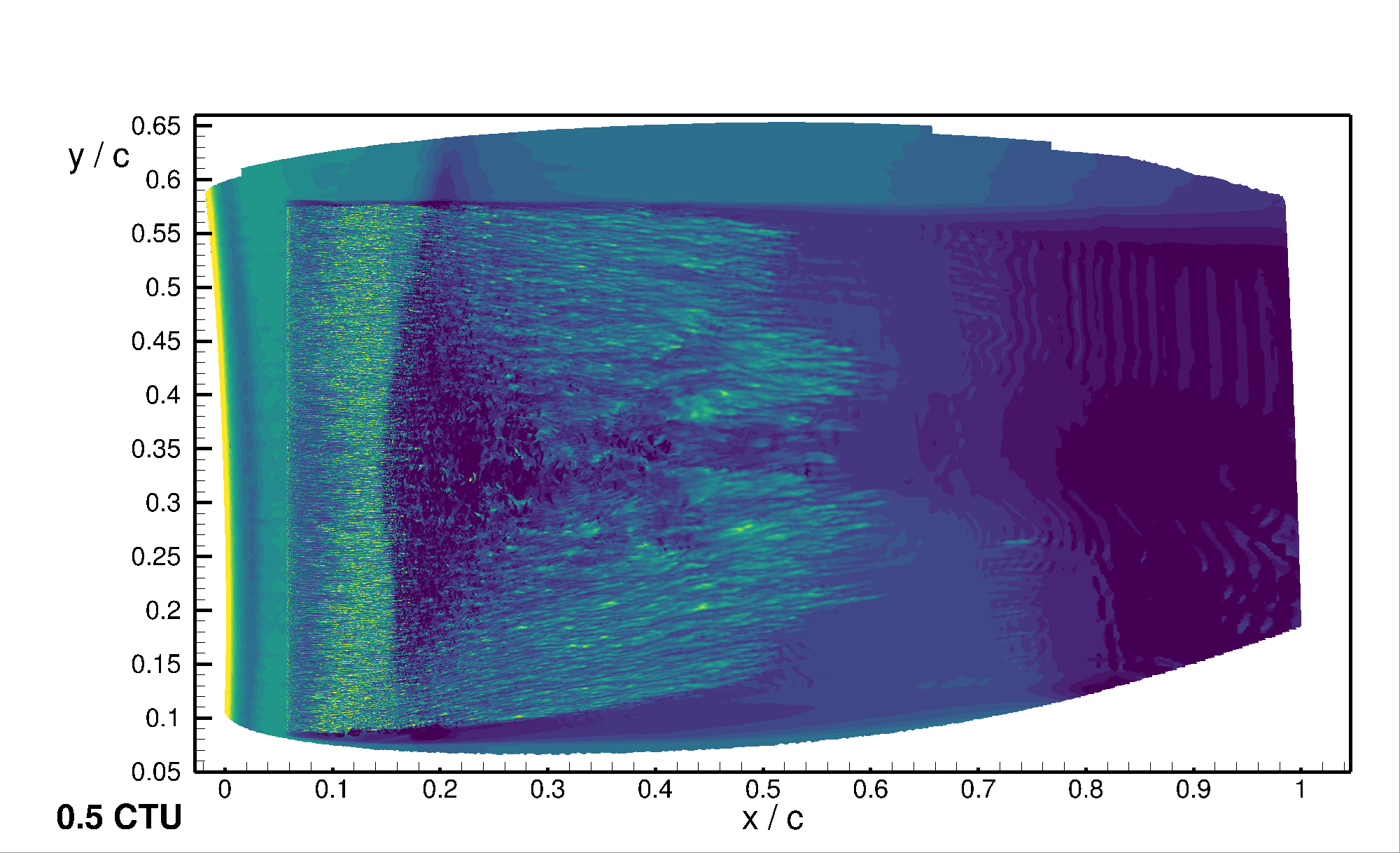}	
	\end{minipage}%
	
	\begin{minipage}[c]{0.6\textwidth}
			\includegraphics[trim= 70 5 60 160,clip,width=\linewidth]{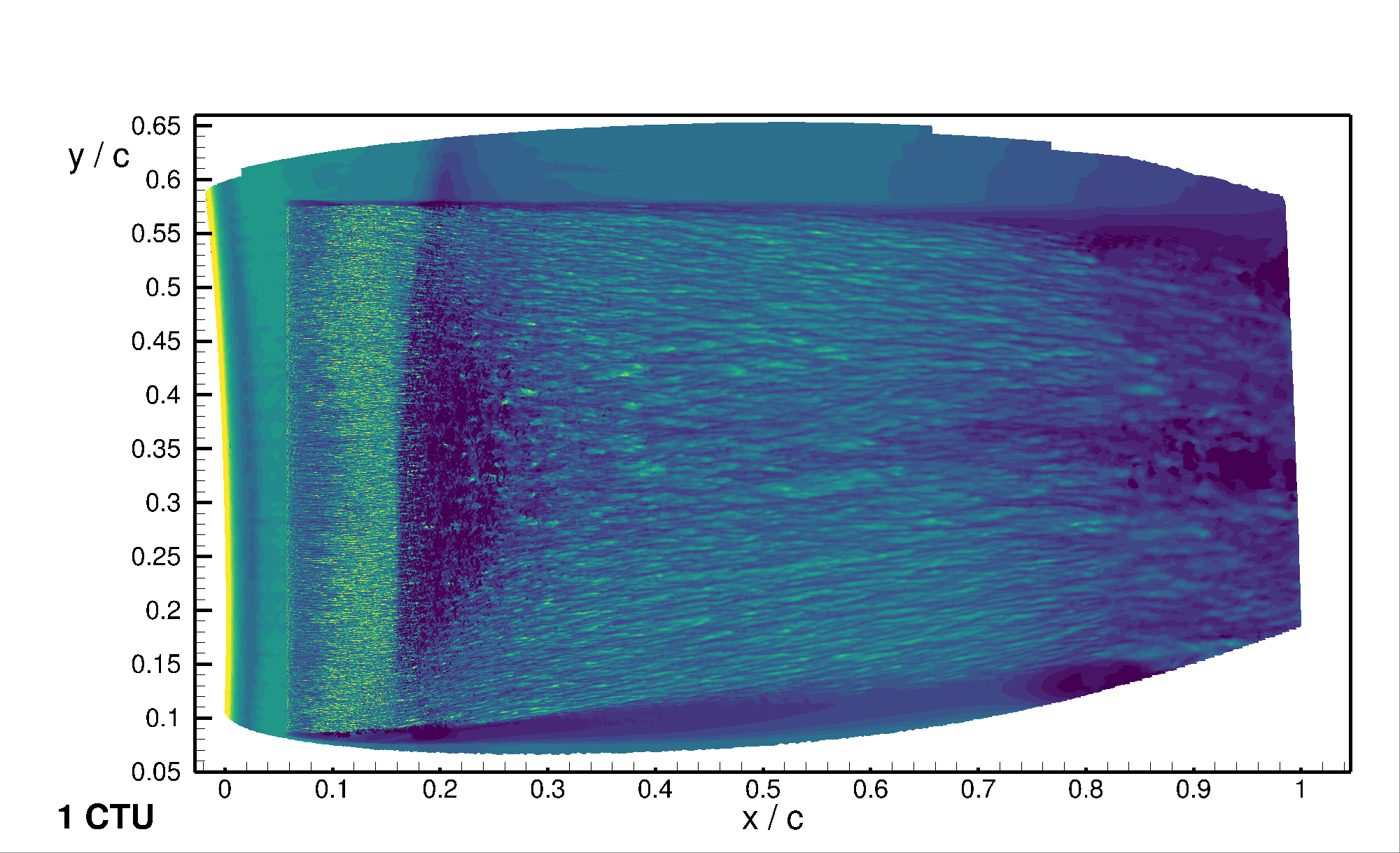}	
	\end{minipage}%
	
	\begin{minipage}[c]{0.6\textwidth}
			\includegraphics[trim= 70 5 60 160,clip,width=\linewidth]{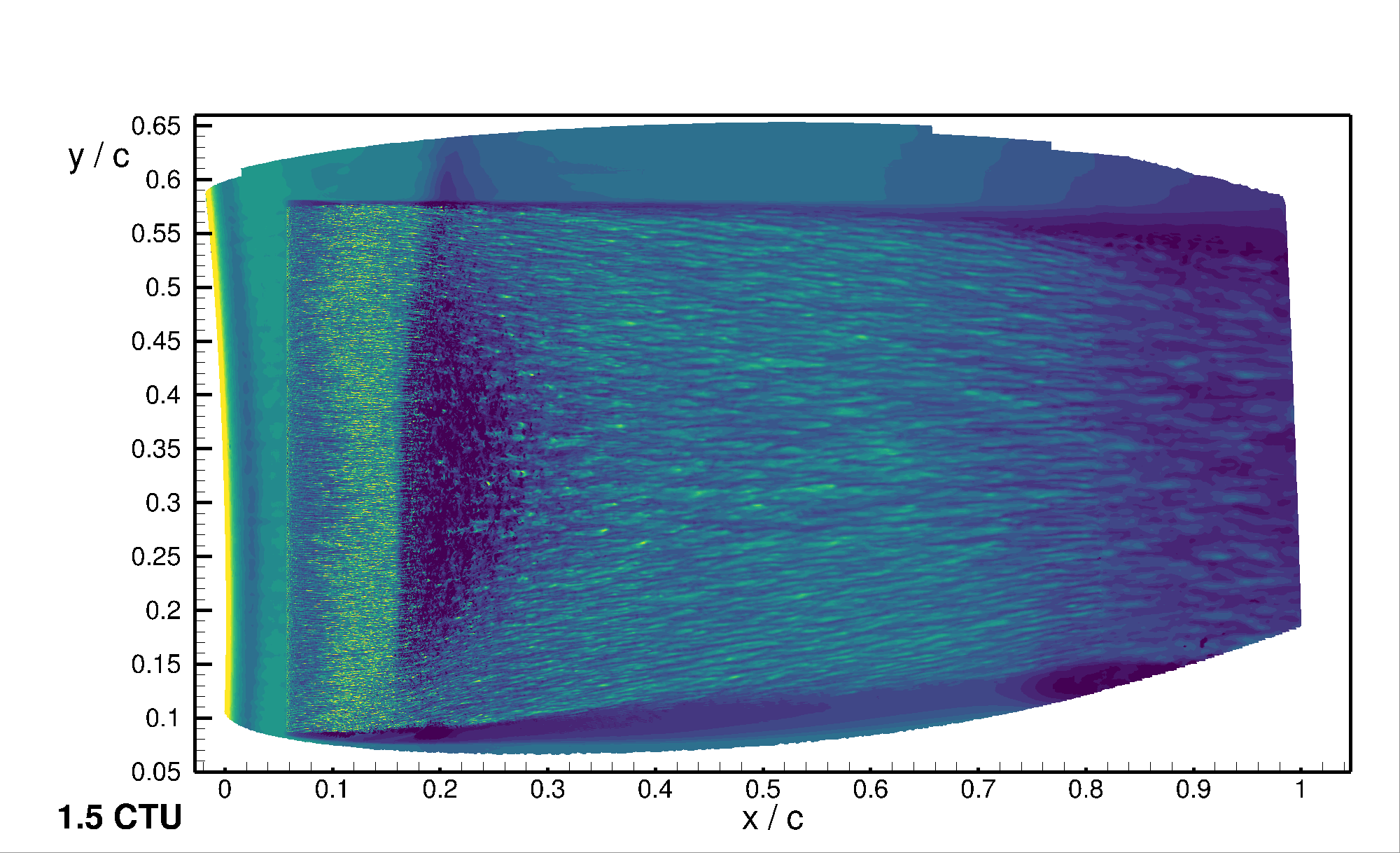}	
	\end{minipage}%

	\caption{Temporal evolution of $c_f$-distribution within the refinement area on projected nacelle surface.}
	\label{abb:snapshot_cf}
	\end{center}
\end{figure}

Another perspective on the temporal evolution is given in Fig. \ref{abb:snapshot_cf}. Here the $c_f$-distribution is shown at four different times. This figure confirms that the resolved turbulence develops over the entire refinement area. The transonic shock front is visible in form of a sudden decrease in $c_f$. As in Fig.  \ref{abb:snapshot_mach} it can be seen that the whole front is moving downstream until it remains in an area of $0.16 \leq x_s/c \leq 0.2$.
A minor numerical effect appears at the lateral edge of the refined mesh in $\varphi$-direction where underresolved turbulence is present.
This is due to the fact that the STG does not directly connect to the lateral RANS zones at the edges of the refinement region.
Therefore two small gaps appear where little resolved and significantly reduced modelled turbulence exists which result in low values of $c_f$. This artefact can easily be circumvented in future simulations by narrowing the LES zone in spanwise direction and thus generate modelled turbulence in the respective regions. Nevertheless, the described phenomenon is limited to the boundaries and does not affect the actual focus region.

\begin{figure}
	\vspace{0cm}
	\begin{center}
	\begin{minipage}[c]{1\textwidth}
			\includegraphics[clip,width=\linewidth]{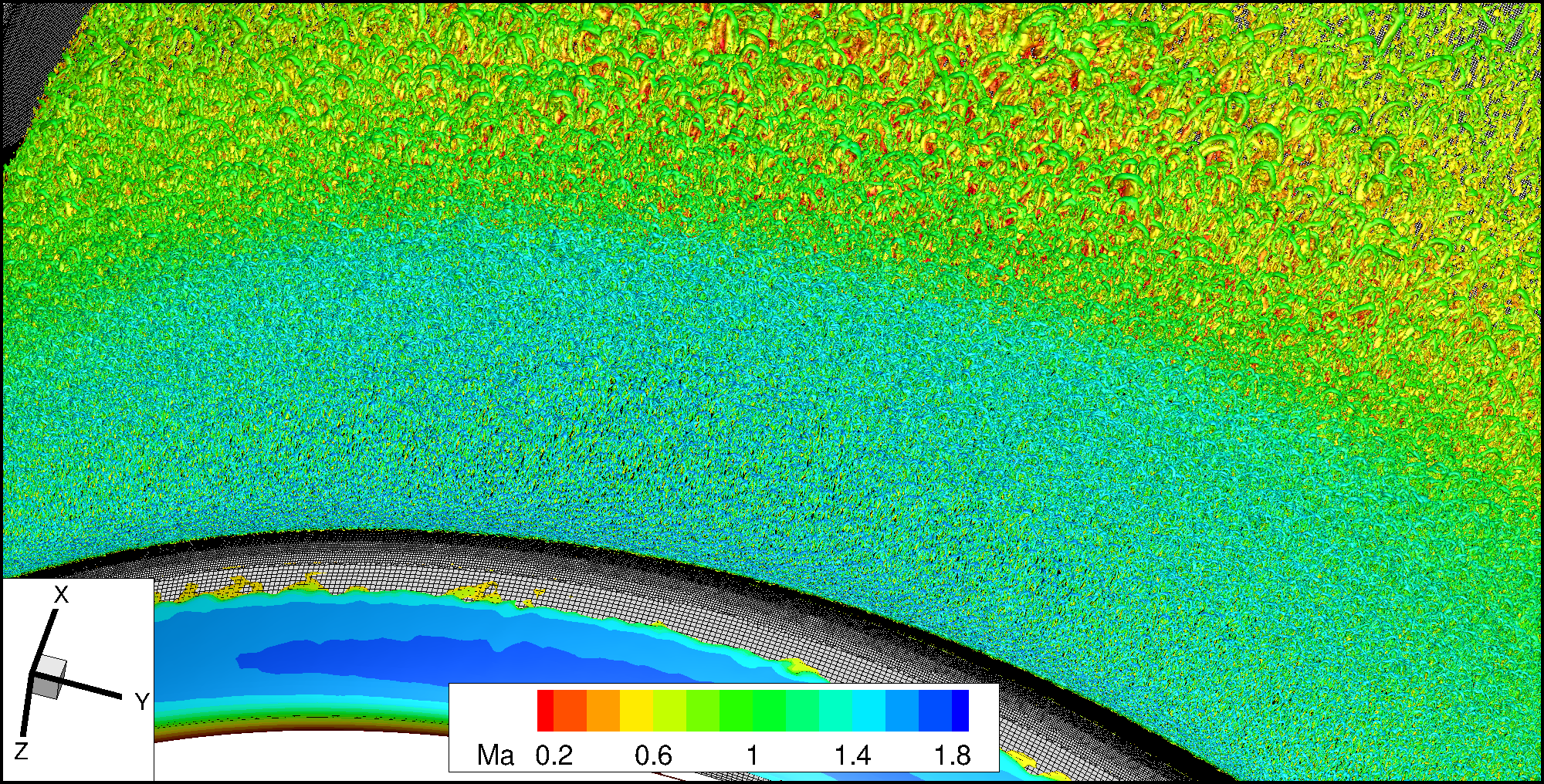}	
	\end{minipage}%

	\caption{Isosurface of Q-Criterion  ($Q=10^{10}$) at nacelle lower surface for LD2 scheme at ${t=1.5\,\text{CTU}}$.}
	\label{abb:snapshot_Q}
	\end{center}
\end{figure}

To give an impression of the vortex structure of the resolved turbulence an isosurface of the $Q$-criterion ($Q=10^{10}$) at $t=1.5$ CTU is depicted in Fig. \ref{abb:snapshot_Q}. As already observed in Fig. \ref{abb:snapshot_cf} an extensive formation of turbulent structures within the refinement region is present. 
These structures are growing with increasing streamwise position and partially evolve into horseshoe vortices which corresponds to expected flow behaviour.

\subsection{Investigation of grey area}
\label{sec:investigation_of_grey_area}

In the following a quantitave analysis of the grey area / adaption region is performed. Therefore the flow field was averaged with regard to time and spanwise direction $\varphi$. The temporal average was applied for $0.42 \leq t/\text{CTU} \leq 1.5$. The start time $t=0.42$ is chosen such that the resolved turbulence is completely established within the focus region ($0.06\leq x/c \leq 0.25$) and no remains of the initial RANS-solution are present in this area (cf. Fig. \ref{abb:snapshot_cf} at $t=0.5\,\text{CTU}$). The spanwise average was applied over the refinement section such that the areas of underresolved turbulence at its margins were omitted ($\varphi \in [125^\circ; 220^\circ]$).

\begin{figure}
	\vspace{0cm}
	\begin{center}
	\begin{minipage}[c]{0.6\textwidth}
			\includegraphics[trim= 100 450 60 150, clip,width=\linewidth]{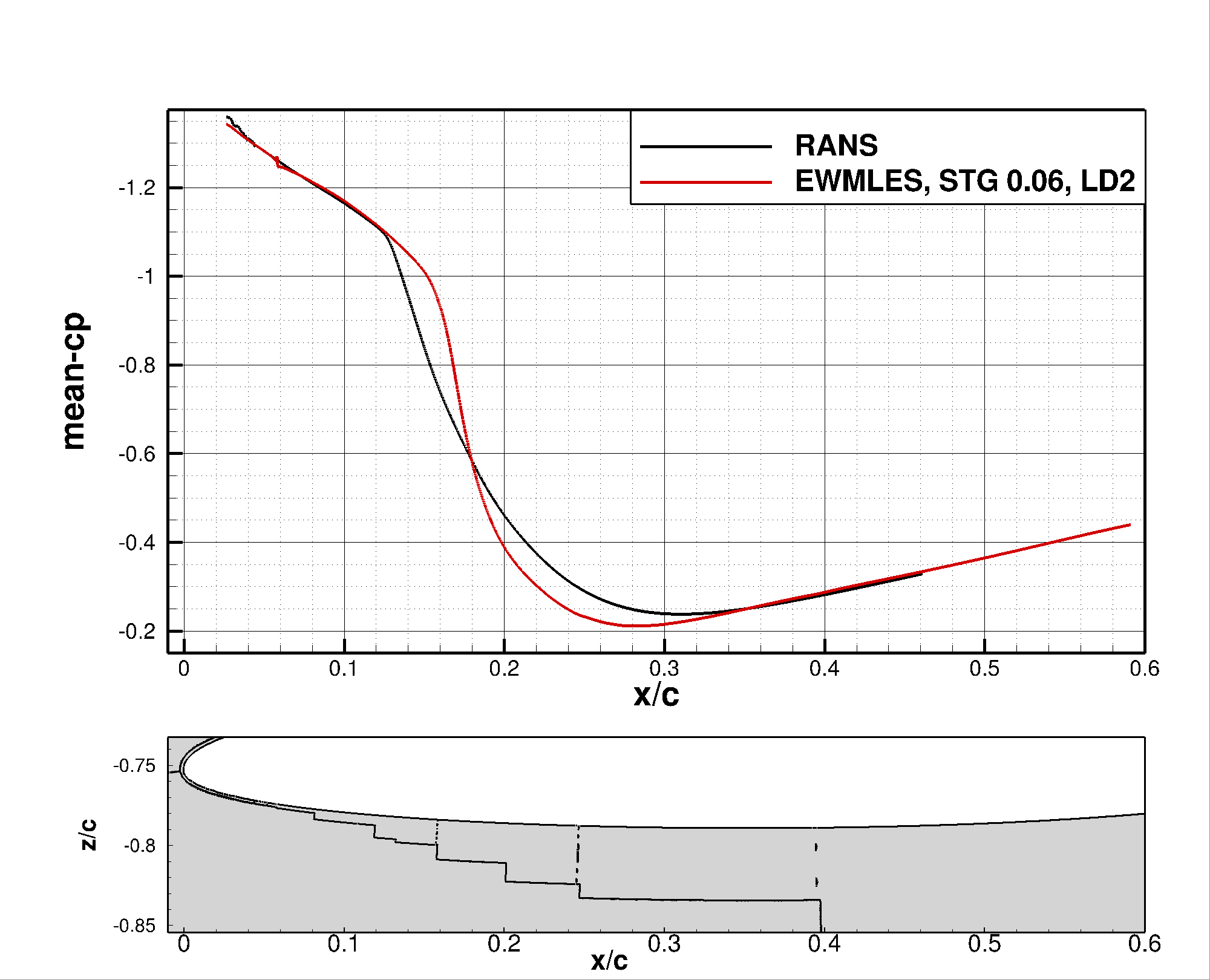}	
	\end{minipage}%
	
	\begin{minipage}[c]{0.6\textwidth}
			\includegraphics[trim= 100 450 60 150, clip,width=\linewidth]{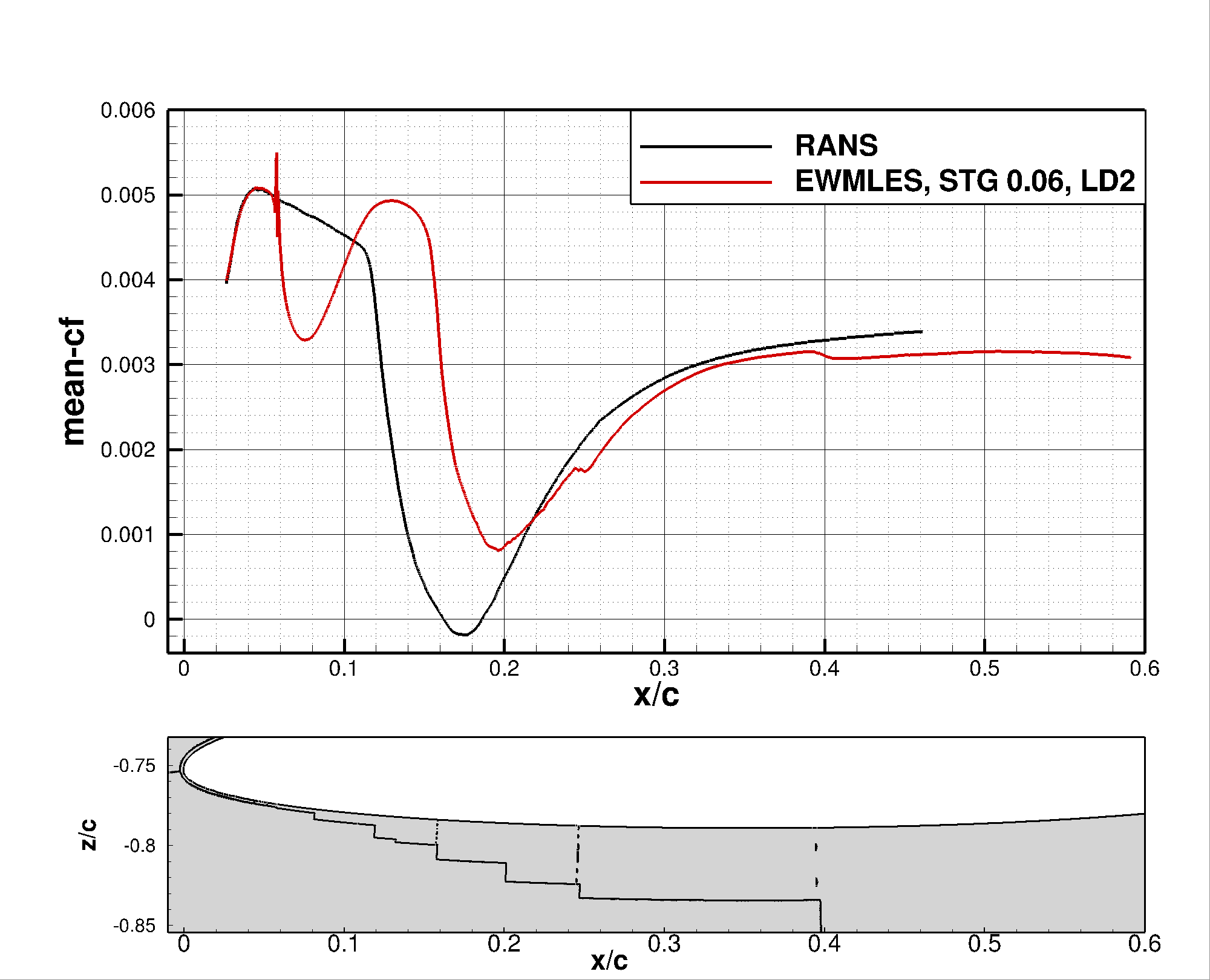}	
	\end{minipage}%

	\caption{Quantitave comparison of time and spanwise averaged pressure - (top) and skin friction distributions (bottom) between the initial RANS and EWMLES solutions.}
	\label{abb:cp_cf_0.06}
	\end{center}
\end{figure}

Fig. \ref{abb:cp_cf_0.06} (top) shows the result of the EWMLES mean pressure distribution (mean-$c_p$) along with the initial RANS solution.
Good agreement between these curves are present for $x/c \leq 0.13$ where $x/c=0.13$ is the average location of the shock front of the SST-RANS solution which results into a sudden rise in mean-$c_p$. It is apparent that this agreement also persists for positions upstream of the STG ($x/c \leq 0.06$) which indicates that no upstream effect of the STG exists.
With regard to the EWMLES shock position the already described shift in downstream direction is also present in this depiction and located at $x/c=0.15$. Due to the comparatively early start in the averaging of mean-$c_p$ it is not reasonable to compare the curves for $x /c \geq 0.3$ since transient effects from the switch from RANS to EWMLES still exist in this area.

A further quantitive flow comparison between SST-RANS and EWMLES is given in Fig. \ref{abb:cp_cf_0.06} (bottom) which shows mean skin friction distributions (mean-$c_f$). 
In the flow region upstream of the STG ($x/c\leq 0.06$) good agreement are visible again which confirms the previously mentioned absence of potential STG upstream effects.
However, for $0.06 \leq x/c \leq 0.16$ remarkable deviations appear. One observes a significant drop in mean-$c_f$ directly downstream of the STG and its increase with a peak value at $x/c=0.13$ and a mean-$c_f$-level which is comparable to the mean-$c_f$ value at the STG position. Although a similar behaviour is present for the flat plate flow as described in Sec. \ref{sec:flat_plate} the flat plate variations in mean-$c_f$ are of significantly smaller.
The adaption length which measures the distance between STG position and subsequent peak in mean-$c_f$ amounts $46\,\delta_{STG}$ where $\delta_{STG}$ represents the boundary layer thickness at the STG position. In case of the flat plate flow this adaption length only amounts $6\,\delta_{STG}$ (cf. Fig. \ref{abb:flat_plate}). A further analysis of these deviations with reference to the flat plate flow are given in Sec. \ref{sec:low_reynolds_number_effect}.
Considering now the region where $0.16\leq x/c \leq 0.25$ we observe that the region of recirculation has disappeared, at least for this transient period of time averaging since mean-$c_f$ is always positive. 
Furthermore additional distortions in the EWMLES mean-$c_f$ distribution appear at $x/c=0.25$ and $x/c=0.40$ which corresponds to locations of the $\Delta \varphi$ coarsening steps of the mesh (cf. Sec. \ref{sec:mesh_resolution}). This indicates that the local mesh resolutions of $r \Delta \varphi = \delta_{\varphi, min } / 10$ might be locally at the lower limit at these positions.

\subsection{Sensitivity studies}
\subsubsection{Positioning of the RANS-LES interface}
\label{sec:results_positioning_STG}
Preliminary grid number estimations for different locations of the RANS-LES interface in $x$-direction ($x_{STG}$) demonstrated a strong dependence of $x_{STG}$ and the total grid number.
A shift of this boundary in downstream direction allows to reduce the total grid number significantly. Exemplarily, moving $x_{STG}$ by $0.02c$ enables to reduce the total grid size about $100\,$Mio points without violating the applied extension and resolution constraints for the refinement area.
This dependence is a consequence of the shortening of the refinement area in $x$-direction by which the subregion with the highest cell density is narrowed. Also, due to the dependence of $\Delta \varphi_{\Omega_1}$ on $\delta_{\varphi, min}(x_{STG})$ in subregion $\Omega_1$ it is possible to increase $\Delta \varphi_{\Omega_1}$ in the entire interval $x/c \in [x_{STG};0.16]$ (cf. \ref{sec:mesh_resolution}).

This dependency on the STG position suggests to place the RANS-LES boundary as close as possible to the shock front and examine its effect on the flow solution. 
Based on the original assumption that the adaption length of the STG amounts less than $10\,\delta_{STG}$ we estimated $x_{STG}/c=0.08$ as latest possible position in order to avoid direct interactions with the shock front. 
Additionally, for this estimation a potential shock movement in upstream direction until ${x_{s, min}=0.1}$ was taken into account. For the following examinations we used the same mesh as before to verify a basic applicability of a late RANS-LES interface.

\begin{figure}
	\vspace{0cm}
	\begin{center}
	\begin{minipage}[c]{0.6\textwidth}
			\includegraphics[trim= 100 450 60 150, clip,width=\linewidth]{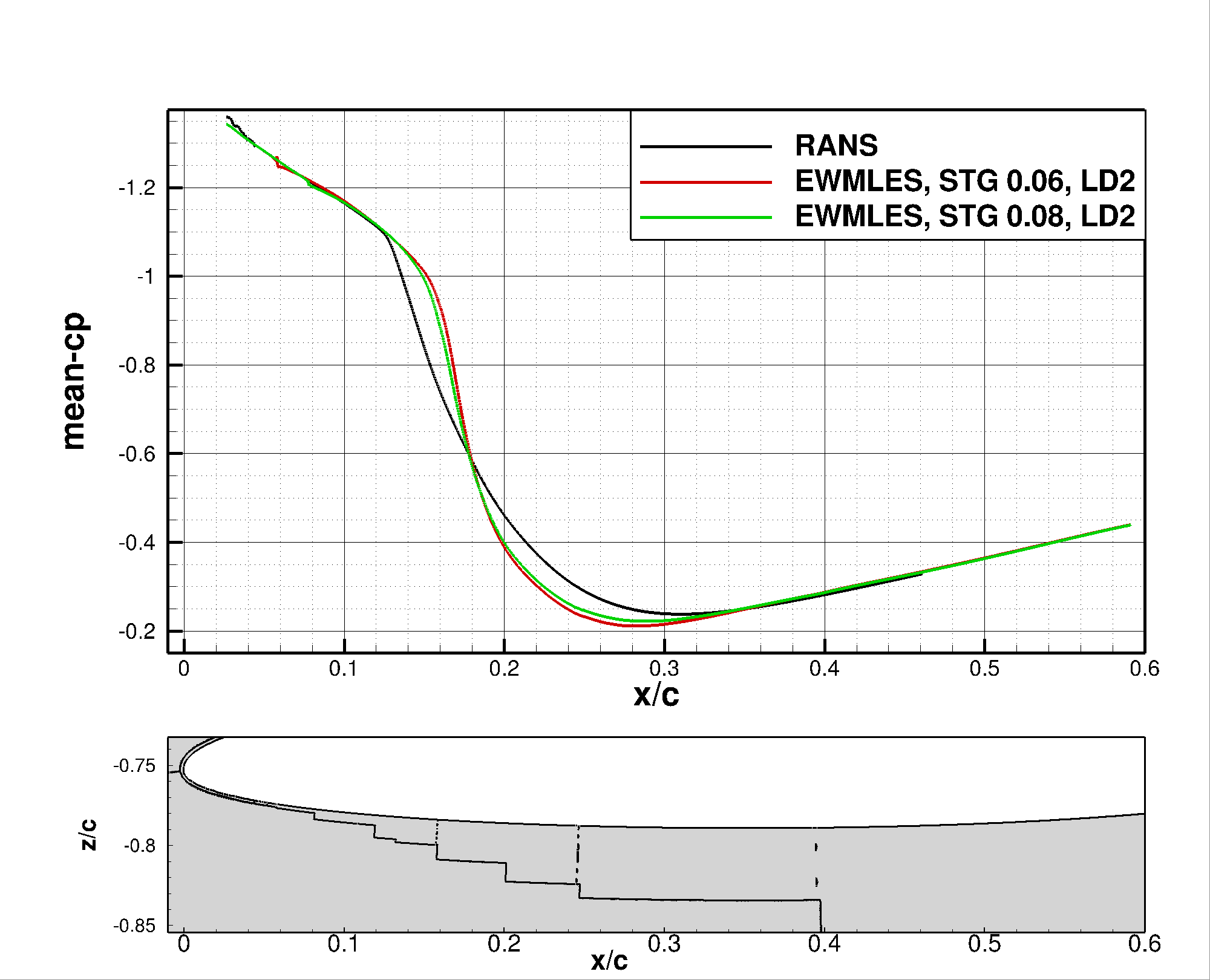}	
	\end{minipage}%
	
	\begin{minipage}[c]{0.6\textwidth}
			\includegraphics[trim= 100 450 60 150, clip,width=\linewidth]{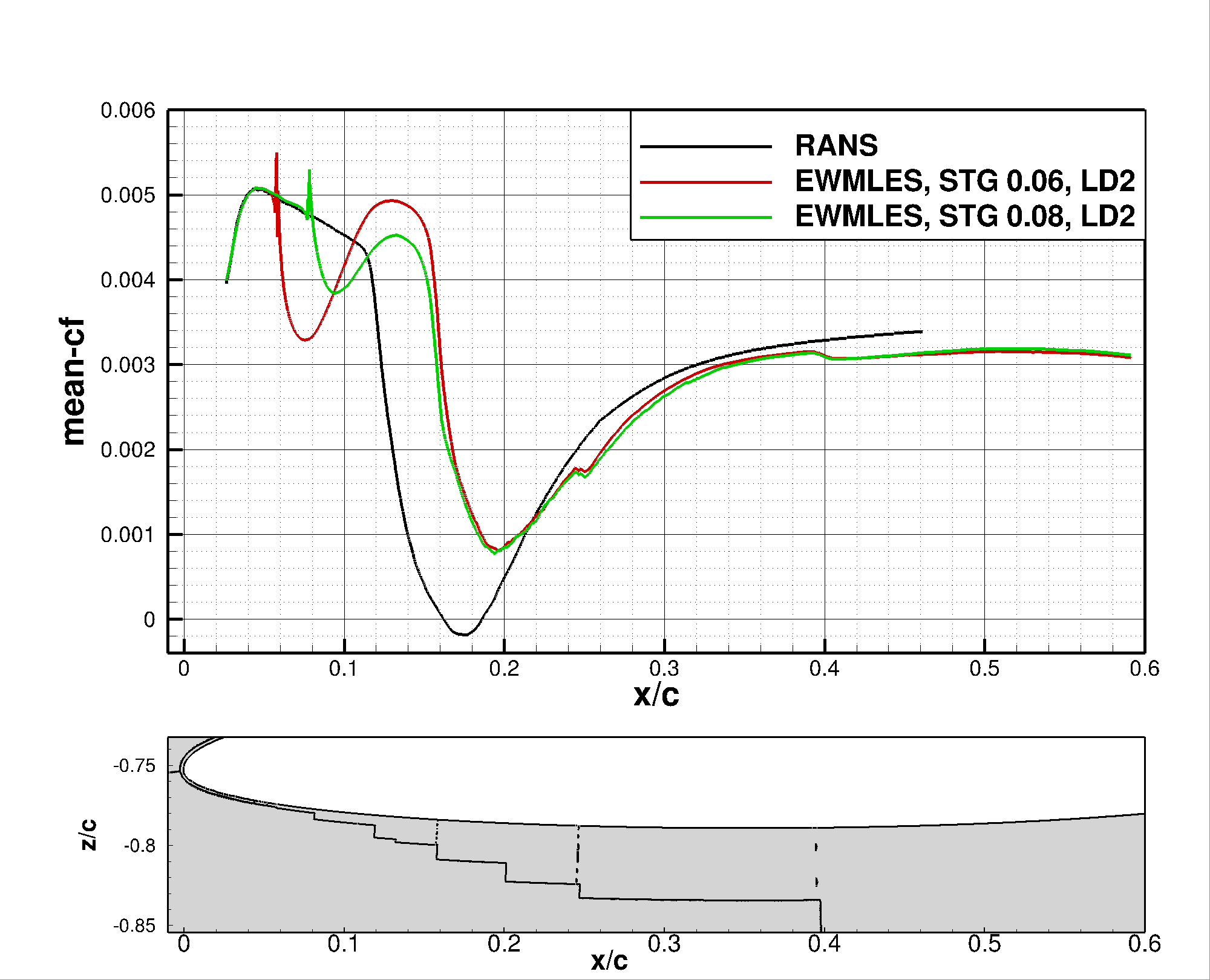}	
	\end{minipage}%

	\caption{Effect of positioning of the RANS-LES interface on averaged surface pressure and skin friction distributions. }
	\label{abb:cp_cf_0.06_0.08}
	\end{center}
\end{figure}

Fig. \ref{abb:cp_cf_0.06_0.08} shows mean-$c_p$ and mean-$c_f$ distributions of the EWMLES results for $x_{STG}/c=0.08$ (green curves) where the same averaging procedure as in Sec. \ref{sec:investigation_of_grey_area} is employed.
It is striking that the mean-$cp$ distribution is almost identical to the previous $x_{STG}/c=0.06$ result (red)
with maximum deviations of two line thicknesses for $x/c\geq 0.16$.
However, with respect to mean-$cf$ and its adaption area downstream of the STG distinct differences compared to the $x_{STG}/c=0.06$ result exist.
Firstly, the initial decay is significantly weaker than before.
Furthermore, its adaption length is reduced and only amounts $19\,\delta_{STG}$ so that its peak is located at almost the same position as for the $x_{STG}/c=0.06$ result.
The peak value though, is significantly reduced and corresponding to the initial RANS solution directly upstream of the shock position.
A further discussion of these features of the adaption regions is given in Sec. \ref{sec:low_reynolds_number_effect}.
It is remarkable that for $x/c \geq 0.16$ the subsequent mean-$c_f$ evolution is almost identical to  the $x_{STG}/c=0.06$ result which demonstrates an independence of the flow solution with regard to the location of the RANS-LES interface.

\subsubsection{Impact of Numerical Scheme}
\label{sec:results_numerical_scheme}
\begin{figure}
	\vspace{0cm}
	\begin{center}
	\begin{minipage}[c]{1\textwidth}
			\includegraphics[clip,width=\linewidth]{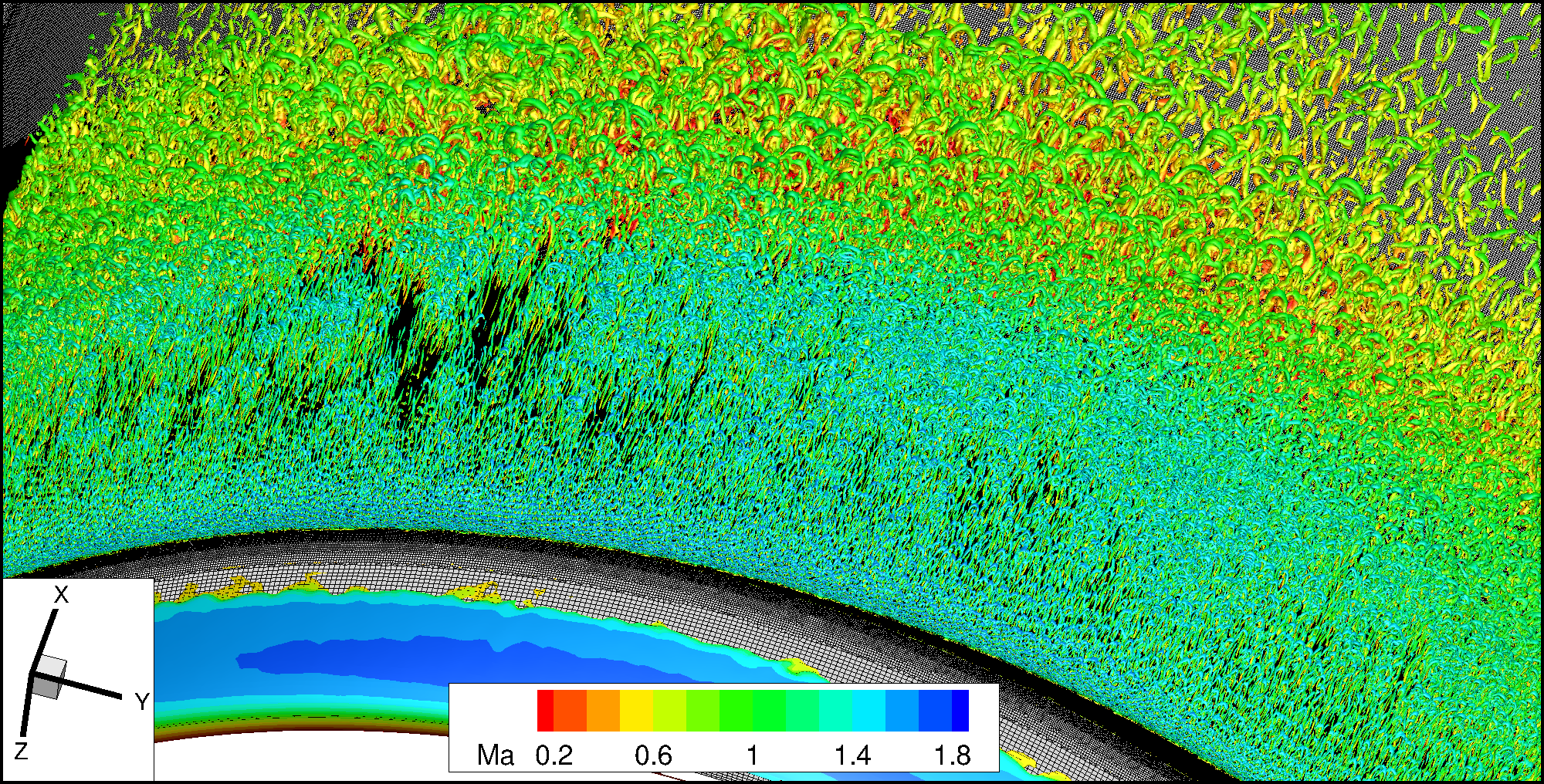}	
	\end{minipage}%

	\caption{Isosurface of Q-Criterion  ($Q=10^{10}$) for reference central-scheme at nacelle lower at ${t=1.5\,\text{CTU}}$.}
	\label{abb:snapshot_Q_reference}
	\end{center}
\end{figure}

A further objective of our research was to compare the effect of different numerical schemes for the central discretisation of viscous fluxes which is applied in the refinement region (LES). 
In addition to the already employed LD2 scheme (Sec. \ref{subsec:LD2}) a reference central-scheme (Eq. \ref{eq:num_blending} in Sec. \ref{subsec:LD2}) is applied on the same numerical setup as in Sec. \ref{sec:investigation_of_grey_area}.
Although the necessity of the high quality LD2 scheme against the reference scheme has been demonstrated with the aid of the DIT-testcase in \ref{sec:dit} it is not obvious how the reference scheme performs for transonic flows on a 3D configuration.
To give a qualitative impression of the flowfield the Q-Criterion at $Q=10^{10}$ for a snapshot at $t=1.5\,\text{CTU}$ is shown in Fig. \ref{abb:snapshot_Q_reference} which can directly compared to Fig. \ref{abb:snapshot_Q}.
The comparison shows that the previous formation of turbulent structures is now partially interrupted. Especially the region directly downstream of the STG lacks turbulent structures. It is striking that coarser structures such as the clearly visible horseshoe vortexes are preserved whereas tiny structures are vanished. This is in direct agreement with the results from the DIT testcase which demonstrates that small turbulent scales are strongly damped by the reference scheme (cf. Fig.\ref{abb:DIT}).

\begin{figure}
	\vspace{0cm}
	\begin{center}
	\begin{minipage}[c]{0.6\textwidth}
			\includegraphics[trim= 100 450 60 150, clip,width=\linewidth]{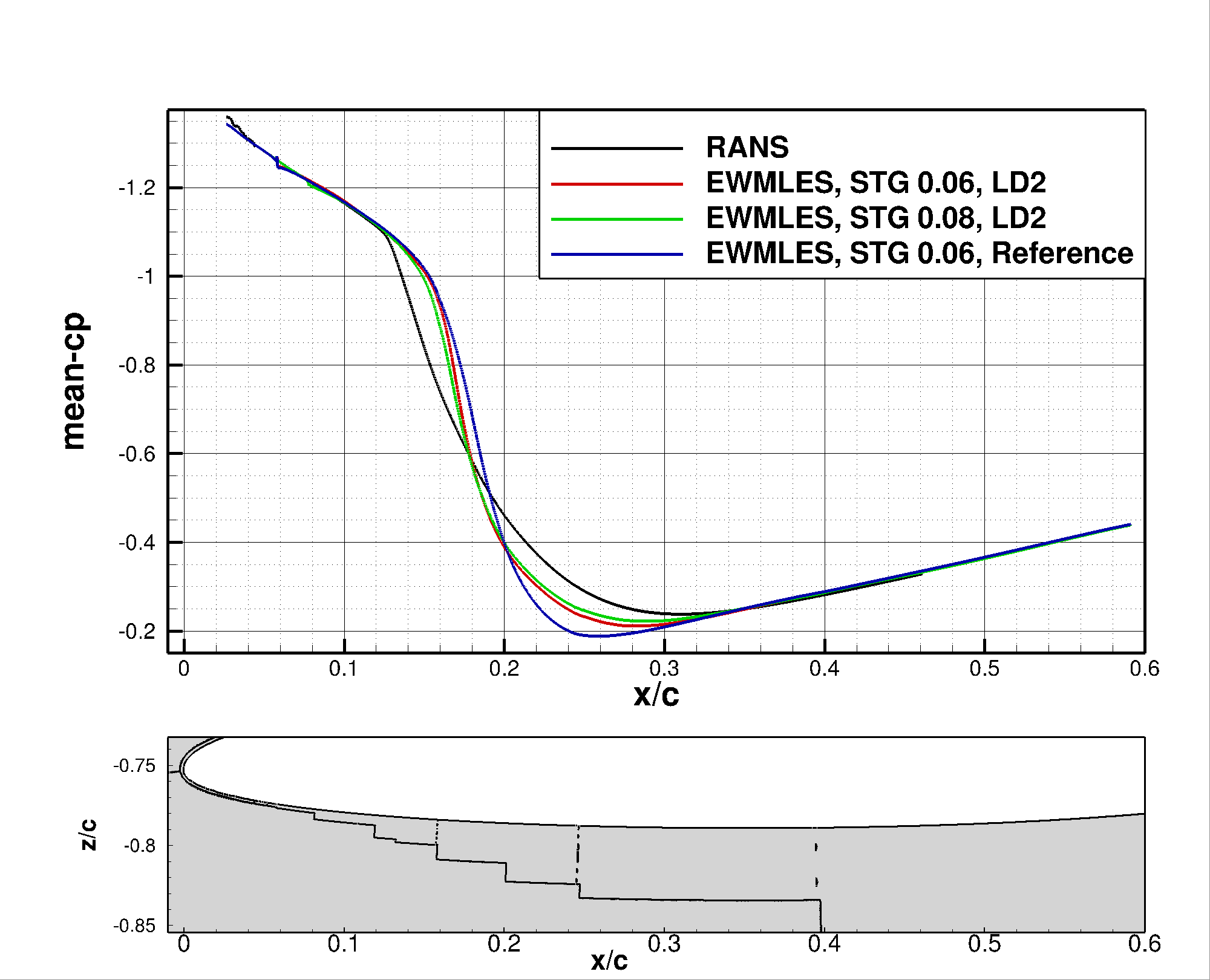}	
	\end{minipage}%
	
	\begin{minipage}[c]{0.6\textwidth}
			\includegraphics[trim= 100 450 60 150, clip,width=\linewidth]{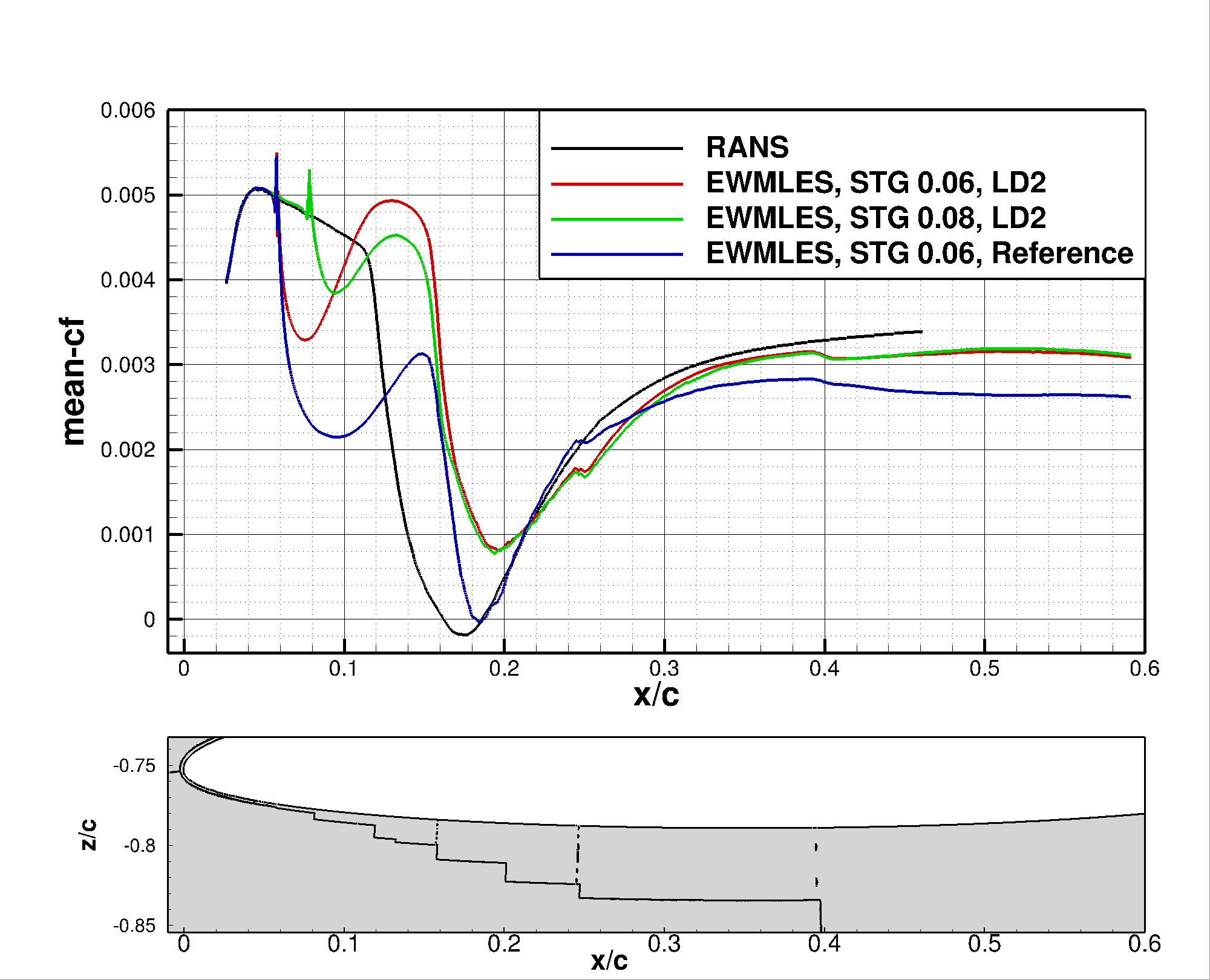}	
	\end{minipage}%

	\caption{Effect of different numerical schemes on averaged surface pressure and skin friction distributions. }
	\label{abb:cp_cf_0.06_0.08_reference}
	\end{center}
\end{figure}

These observations are also present in the analysis of the average skin friction distribution (blue curve in Fig. \ref{abb:cp_cf_0.06_0.08_reference}). 
Whereas the mean surface pressure is hardly affected by the numerical scheme, mean-$c_f$ shows large deviations. Especially the decay downstream of the STG indicates a lack of resolved turbulence.
Additionally, compared to the LD2 results the mean-$c_f$ level is underestimated 
in the area downstream of the shock - boundary layer interaction ($0.35\leq x/c\leq0.6$). This confirms the previous observation of Fig. \ref{abb:snapshot_Q_reference} of underresolved turbulence throughout the entire refinement region.

\subsection{Reynolds number and mesh resolution effect on STG adaption region}
\label{sec:low_reynolds_number_effect}
In the following we address the so far unsound behaviour of the adaption region downstream of the STG arising for all shown configurations.
As already described before the adaption region displays the largest deviations with regard to adaption length as well as maximal and minimal mean-$c_f$-deviations for the nacelle at $x_{STG}=0.06c$. These features reduce for $x_{STG}=0.08c$ and almost vanish but are still present for the flat plate test case (cf. Fig. \ref{abb:flat_plate} and \ref{abb:cp_cf_0.06_0.08}).
A closer look into the flow properties and mesh resolution at the location of the STG suggests a dependency on $Re_{\delta, STG}$ (Tab. \ref{tab:reynolds_number}).
Here, $Re_{\delta,  STG}$ is defined as a Reynolds number referring to the local boundary layer thickness $\delta_{STG}$ as well as velocity and kinematic viscosity at the outer edge of $\delta_{STG}$. 
This Reynolds number, which directly impacts the input statistics of the STG, has its lowest number for the nacelle case at $x_{STG}=0.06c$ (4989) and increases for $x_{STG}=0.06c$ (6975) and the flat plate flow (24200). 
The ratio of turbulent- and laminar viscosity ($\max \left(  \mu_t / \mu_l \right) $) which serves as measure of modelled turbulence shows a comparable trend.
Since low Reynolds numbers enhance the stability of the boundary layer and hence suppress turbulent fluctuations, this might lead to a damping of the injected turbulent structures. As a consequence the boundary layer evolves into a flow with significantly reduced turbulence which is visible in a strongly reduced level of mean-$cf$. Thus, it appears that the distinct adaption region can be traced back to a low-Reynolds number effect.

Another reason might be due to the mesh resolution $\Delta y$ which amounts $\delta/20$ for the flat plate flow and coarsens to $\delta/16$ and $\delta / 12$ for $x_{STG}=0.08c$ and $x_{STG}=0.06c$, respectively (cf. Tab. \ref{tab:reynolds_number}). Since a resolution of $\Delta y =\delta/20$ is actually defined as coarsest resolution in this flow direction the here observed somewhat
coarser resolutions might perturb a proper development of the turbulent boundary layer \cite{shur2008hybrid}.

Therefore further examinations of the transonic nacelle flow for higher $Re_\infty$ (resulting in larger $Re_\delta$) as well as finer resolutions $\Delta y$ will be performed in future work in order to provide a verification of the here detected limits of synthetic turbulence generation at locally low Reynolds numbers.

\begin{table}[ht]
\begin{center}
\begingroup
\renewcommand{\arraystretch}{1.5} 
\begin{tabular}
{l| l| l| l| l| l| l}
& $Re_{\infty}$   &  $\delta_{STG}$/m  &  $Re_{\delta,  STG}$ & $\Delta   x$ & $\Delta y   $ &  $\max \left(  \mu_t / \mu_l \right) $ \\ \hline \hline

Flat Plate        & 4.7 Mio & 0.006                                  & 24200                                            & $\delta $/10         & $\delta $/20          & 87                                            \\  

      &      &      &    &      &       \\  \hline
Nacelle   & 3.3 Mio & 0.00024                                & 4989                                             & $\delta $/11.2       & $\delta $/11.76       & 9                                             \\ 

 $x_{STG}=0.06c$ &  &           &             &      &     &   \\ \hline

Nacelle & 3.3 Mio & 0.00033                                & 6975                                             & $\delta $/13.75      & $\delta $/16.17       & 10        \\

 $x_{STG}=0.08c$ &  &           &             &      &     &   \\ 
 
\end{tabular}
\caption{Comparison of several local flow quantities at the location of the synthetic turbulence generator for all presented configurations.}
\label{tab:reynolds_number}
\endgroup
\end{center}
\end{table}

\section{Conclusions}\label{sec13}
\label{sec:conclusions}
A scale-resolving WMLES methodology in conjunction with the SST turbulence model was applied to the XRF-1 aircraft configuration with UHBR nacelle at transonic flow conditions. The method was applied locally at the nacelle surface in order to examine shock induced separation. A Synthetic Turbulence Generator (STG) was employed to enhance the transition from modelled to resolved turbulence at the RANS-LES interface.

Prior to the actual examination on the aircraft configurations basic functionalities of the methodology were successfully verified for flows of decaying isotropic turbulence and a flow over a flat plate for $Re_{\theta}=3030$. 

With regard to the target configuration a sophisticated mesh which refines $32\,\%$ of the nacelle outer surfaces and comprises 420 million grid points was constructed.
The main features of the mesh design are the dependence of mesh resolution ($\Delta x, \Delta y$ and $\Delta z$) on the local boundary layer thickness and the consideration of a potential shock movement due to buffet.

Analysis of the transient process of the simulation showed a well resolved formation of turbulent structures over almost the entire refinement region with a broad spectrum of turbulent scales. 
It has been demonstrated that these features are also the result of the employed LD2 scheme. For a reference central-scheme with higher artificial dissipation, small turbulent scales are damped leading to globally underresolved turbulence.

Another outcome of this study is the observation that the STG - adaption region correlates to the local Reynolds number as well as mesh resolution in spanwise direction. For decreasing Reynolds numbers and coarser mesh resolutions an increasing adaption length and more distinct decay in the skin friction distribution were observed.
We note that the methodology is only applicable if the STG adaption region does not interfere with the transonic shock front and therefore sufficient distance to the shock is required.
This distance might not be given in case of an upstream moving shock which would arise for strong shock buffet at the given Reynolds number. 
Therefore further research on the transonic nacelle flow for higher Reynolds numbers as well as finer resolutions will be performed in future work to verify a potential reduction of the adaption length.

\backmatter

\bmhead{Acknowledgments}
The authors gratefully acknowledge the Deutsche 
Forschungsgemeinschaft DFG (German Research Foundation) for funding 
this work in the framework of the research unit FOR 2895.
The authors thank the Helmholtz Gemeinschaft HGF (Helmholtz Association),
Deutsches Zentrum für Luft- und Raumfahrt DLR (German AerospaceCenter) and Airbus for providing the wind tunnel model and financing the wind tunnel measurements
Additionally, the authors gratefully acknowledge the computing time granted by the Resource Allocation Board and provided on the supercomputer Lise and Emmy at NHR@ZIB and NHR@Göttingen as part of the NHR infrastructure. The calculations for this research were conducted with computing resources under the project nii00164.

\section*{Declarations}
\begin{itemize}
\item Funding: This study was funded by DFG (German Research Foundation).
\item Competing interests: The authors have no competing interests to declare that are relevant to the content of this article.
\item Ethics approval: Not applicable
\item Consent to participate: Not applicable
\item Consent for publication: Not applicable
\item Availability of data and materials: Not applicable
\item Code availability: Not applicable
\item Authors' contributions: Not applicable
\end{itemize}

\bibliography{sn-bibliography}%

\begin{thebibliography}{10}
\providecommand{\url}[1]{{#1}}
\providecommand{\urlprefix}{URL }
\providecommand{\doi}[1]{\url{https://doi.org/#1}}
\bibcommenthead

\bibitem{spinner2021design}
S.~Spinner, R.~Rudnik, Design of a uhbr through flow nacelle for high speed
  stall wind tunnel investigations.
\newblock Deutscher Luft- und Raumfahrt Kongress  (2021)

\bibitem{cecora2015differential}
R.D. C{\'e}cora, R.~Radespiel, B.~Eisfeld, A.~Probst, Differential
  reynolds-stress modeling for aeronautics.
\newblock AIAA Journal \textbf{53}(3), 739--755 (2015)

\bibitem{shur2008hybrid}
M.L. Shur, P.R. Spalart, M.K. Strelets, A.K. Travin, A hybrid rans-les approach
  with delayed-des and wall-modelled les capabilities.
\newblock International journal of heat and fluid flow \textbf{29}(6),
  1638--1649 (2008)

\bibitem{Travin2002}
A.~Travin, M.~Shur, M.~Strelets, P.R. Spalart, {Physical and Numerical Upgrades
  in the Detached-Eddy Simulation of Complex Turbulent Flows}.
\newblock Advances in LES of Complex Flows \textbf{65}(5), 239--254 (2002)

\bibitem{Schwamborn2006}
D.~Schwamborn, T.~Gerhold, R.~Heinrich, in \emph{ECCOMAS CFD, P. Wesseling, E.
  O{\~{n}}ate, J. P{\'{e}}riaux (Eds), TU Delft, The Netherlands}, ed. by
  M.~Braza, A.~Bottaro, M.~Thompson (2006)

\bibitem{Menter1994}
F.R. Menter, {Two-Equation Eddy-Viscosity Turbulence Models for Engineering
  Applications}.
\newblock AIAA journal \textbf{32}(8), 1598--1605 (1994)

\bibitem{probst2017evaluation}
A.~Probst, D.~Schwamborn, A.~Garbaruk, E.~Guseva, M.~Shur, M.~Strelets,
  A.~Travin, Evaluation of grey area mitigation tools within zonal and
  non-zonal rans-les approaches in flows with pressure induced separation.
\newblock International Journal of Heat and Fluid Flow \textbf{68}, 237--247
  (2017)

\bibitem{Adamian2011a}
D.~Adamian, A.~Travin, in \emph{Computational Fluid Dynamics 2010}, ed. by
  A.~Kuzmin (Springer Berlin Heidelberg, 2011), pp. 739--744.
\newblock \doi{10.1007/978-3-642-17884-9}

\bibitem{francois2015forced}
D.G. Francois, R.~Radespiel, A.~Probst, {Forced synthetic turbulence approach
  to stimulate resolved turbulence generation in embedded LES}.
\newblock Notes on Numerical Fluid Mechanics and Multidisciplinary Design
  \textbf{130}, 81--92 (2015).
\newblock \doi{10.1007/978-3-319-15141-0_6}

\bibitem{Probst2020}
A.~Probst, P.~Str{\"{o}}er, {Comparative Assessment of Synthetic Turbulence
  Methods in an Unstructured Compressible Flow Solver}.
\newblock Notes on Numerical Fluid Mechanics and Multidisciplinary Design
  \textbf{143}, 193--202 (2020).
\newblock \doi{10.1007/978-3-030-27607-2_15}

\bibitem{Probst2016}
A.~Probst, J.~L{\"{o}}we, S.~Reu{\ss}, T.~Knopp, R.~Kessler, {Scale-Resolving
  Simulations with a Low-Dissipation Low-Dispersion Second-Order Scheme for
  Unstructured Flow Solvers}.
\newblock AIAA Journal \textbf{54}(10), 2972--2987 (2016)

\bibitem{kok2009high}
J.~Kok, A high-order low-dispersion symmetry-preserving finite-volume method
  for compressible flow on curvilinear grids.
\newblock Journal of Computational Physics \textbf{228}(18), 6811--6832 (2009)

\bibitem{Loewe2016}
J.~L{\"{o}}we, A.~Probst, T.~Knopp, R.~Kessler, {Low-Dissipation Low-Dispersion
  Second-Order Scheme for Unstructured Finite-Volume Flow Solvers}.
\newblock AIAA Journal \textbf{54}(10), 2961--2971 (2016)

\bibitem{Probst2022}
A.~Probst, S.~Melber-Wilkending, {Hybrid RANS/LES of a generic high-lift
  aircraft configuration near maximum lift}.
\newblock International Journal of Numerical Methods for Heat {\&} Fluid Flow
  \textbf{32}(4), 1204--1221 (2022).
\newblock \doi{10.1108/hff-08-2021-0525}

\bibitem{comte1971simple}
G.~Comte-Bellot, S.~Corrsin, Simple eulerian time correlation of full-and
  narrow-band velocity signals in grid-generated,‘isotropic’turbulence.
\newblock Journal of fluid mechanics \textbf{48}(2), 273--337 (1971)

\bibitem{Kraichnan1970}
R.H. Kraichnan, {Diffusion by a Random Velocity Field}.
\newblock The Physics of Fluids \textbf{13}(1), 22--31 (1970)

\bibitem{Probst2018}
A.~Probst, {Implementation and assessment of the synthetic-eddy method in an
  unstructured compressible flow solver}.
\newblock Notes on Numerical Fluid Mechanics and Multidisciplinary Design
  \textbf{137}, 91--101 (2018).
\newblock \doi{10.1007/978-3-319-70031-1_7}

\bibitem{Laraufie2013}
R.~Laraufie, S.~Deck, {Assessment of Reynolds stresses tensor reconstruction
  methods for synthetic turbulent inflow conditions. Application to hybrid
  RANS/LES methods}.
\newblock International Journal of Heat and Fluid Flow \textbf{42}, 68--78
  (2013).
\newblock \doi{10.1016/j.ijheatfluidflow.2013.04.007}

\bibitem{nagib2007approach}
H.M. Nagib, K.A. Chauhan, P.A. Monkewitz, Approach to an asymptotic state for
  zero pressure gradient turbulent boundary layers.
\newblock Philosophical Transactions of the Royal Society A: Mathematical,
  Physical and Engineering Sciences \textbf{365}(1852), 755--770 (2007)

\bibitem{franccois2020development}
D.G. Fran{\c{c}}ois, \emph{Development of an Efficient Synthetic Turbulence
  Generator for Hybrid RANS/LES Methods} (TU Braunschweig-Nieders{\"a}chsisches
  Forschungszentrum f{\"u}r Luftfahrt, 2020)

\bibitem{spalart2001young}
P.R. Spalart, C.~Streett, Young-person's guide to detached-eddy simulation
  grids.
\newblock NASA Technical Reports Server  (2001)

\bibitem{menter2012best}
F.R. Menter, Best practice: scale-resolving simulations in ansys cfd.
\newblock ANSYS Germany GmbH \textbf{1} (2012)

\bibitem{jacquin2009experimental}
L.~Jacquin, P.~Molton, S.~Deck, B.~Maury, D.~Soulevant, Experimental study of
  shock oscillation over a transonic supercritical profile.
\newblock AIAA journal \textbf{47}(9), 1985--1994 (2009)

\end{thebibliography}

\end{document}